\documentclass[aps,preview,showpacs,preprintnumbers,twocolumn,amsmath,amssymb,superscriptaddress,nofootinbib,longbibliography]{revtex4-1}

\usepackage{graphicx}
\usepackage{bm}
\usepackage[usenames,dvipsnames]{xcolor}
\usepackage{hyperref}
\usepackage{pstricks}
\usepackage[tight]{subfigure}
\usepackage{verbatim}
\usepackage{units}
\usepackage{multirow}
\usepackage{enumitem}
\usepackage[rightcaption]{sidecap}
\usepackage{mathrsfs}

\newcommand{\kFw}{k_\text{F}^\text{w}}
\newcommand{\vFw}{v_\text{F}^\text{w}}
\newcommand{\EFw}{E_\text{F}^\text{w}}
\newcommand{\kw}{k^\text{w}}
\newcommand{\Eg}{E_\text{g}}
\newcommand{\EF}{E_\text{F}}
\newcommand{\me}{m_\text{e}}

\definecolor{CM}{rgb}{0.2,0.0,1.0} 

\begin{document}

\title[Impact of tunnel barrier strength of magnetoresistance in CNTs]
{Impact of tunnel barrier strength on magnetoresistance in carbon nanotubes}

\author{Caitlin Morgan}
	\affiliation{Peter Gr\"unberg Institut (PGI-6), Forschungszentrum J\"ulich, D-52425 J\"ulich, Germany}
	\affiliation{JARA--Fundamentals of Future Information Technologies}
\author{Maciej Misiorny}
    \affiliation{Department of Microtechnology and Nanoscience MC2, Chalmers University of Technology, SE-412 96 G\"{o}teborg, Sweden}
    \affiliation{Peter Gr{\"u}nberg Institut (PGI-2), Forschungszentrum J{\"u}lich, D-52 425 J{\"u}lich,  Germany}
    \affiliation{Faculty of Physics, Adam Mickiewicz University, PL-61 614 Pozna\'{n}, Poland}
\author{Dominik Metten}
	\affiliation{Institut de Physique et Chimie des Mat\'eriaux de Strasbourg and NIE, UMR 7504, Universit\'e de Strasbourg and CNRS, France}
\author{Sebastian Heedt}
	\affiliation{Peter Gr\"unberg Institut (PGI-9), Forschungszentrum J\"ulich, D-52425 J\"ulich, Germany}
	\affiliation{JARA--Fundamentals of Future Information Technologies}
\author{Thomas Sch\"apers}
	\affiliation{Peter Gr\"unberg Institut (PGI-9), Forschungszentrum J\"ulich, D-52425 J\"ulich, Germany}
	\affiliation{JARA--Fundamentals of Future Information Technologies}
\author{Claus M. Schneider}
	\affiliation{Peter Gr\"unberg Institut (PGI-6), Forschungszentrum J\"ulich, D-52425 J\"ulich, Germany}
	\affiliation{JARA--Fundamentals of Future Information Technologies}
	\affiliation{Fakult\"at f.~Physik and Center for Nanointegration Duisburg-Essen (CeNIDE), Universit\"at Duisburg-Essen, D-47048 Duisburg, Germany }
\author{Carola Meyer}\email{carola.meyer@uos.de}
	\affiliation{Peter Gr\"unberg Institut (PGI-6), Forschungszentrum J\"ulich, D-52425 J\"ulich, Germany}
	\affiliation{JARA--Fundamentals of Future Information Technologies}
	\affiliation{Fachbereich Physik, Universit\"at Osnabr\"uck, D-49069 Osnabr\"uck, Germany}

\date{\today}

\begin{abstract}
We investigate magnetoresistance in spin valves involving CoPd-contacted carbon nanotubes. Both temperature and bias voltage dependence clearly indicate tunneling magnetoresistance as the origin. We show that this effect is significantly affected by	the tunnel barrier strength, which appears to be one reason for the variation between devices previously detected in similar structures. Modeling the data by means of the scattering matrix approach, we find a non-trivial dependence of  the magnetoresistance on the barrier strength. Furthermore, analysis of the spin precession observed in a nonlocal Hanle measurement yields a spin lifetime of $\tau_\text{s} =1.1$ ns, a value comparable with those found in silicon- or graphene-based spin valve devices.
\end{abstract}

\maketitle

\section{Introduction}
Low dimensional carbon systems, such as graphene and carbon nanotubes (CNTs), are envisaged as promising candidates for electronic devices beyond the conventional CMOS technology. In the field of spintronics~\cite{Zutic04}, considerable interest in such systems stems from their inherent properties: long coherence times manifesting as ballistic transport of electrons, and long spin lifetimes due to low spin-orbit coupling and hyperfine interaction that can be even zero, if $^{12}$C is used in the synthesis. Spin injection into two-dimensional graphene is a well-understood phenomenon, and it has been experimentally established that reliable contacts are formed with MgO or Al$_2$O$_3$ as the tunnel barrier~\cite{Han14, Droegerler14, Dlubak12}.  {However, there was a strong preceding debate about the role of the insulator with regard to spin injection~\cite{Tombros07, Han10}, and ongoing research strives to improve the tunnel barrier even further~\cite{Kamalakar14}.}

Understanding the process of spin injection in its one-dimensional counterpart has proven to be far more difficult, mainly due to two reasons. First, it is difficult to reliably fabricate low-resistive ferromagnetic contacts to CNTs. As a result, a substantial variation of the magnitude of the magnetoresistance~(MR) effect between different devices has been observed from early on~\cite{Ago99}. Second, the underlying physics of the MR in CNTs is more complex, since the transport regime is determined by the strength of the tunnel coupling between a CNT and the leads. The two limiting cases are the quantum dot regime, when the coupling is \emph{weak}, and the Fabry-P\'erot regime for \emph{strong} coupling leading to high transmission~\cite{Mann03}. In the former case, huge variations of the MR effect between  {$-$80\% and +120\%  have been observed~\cite{Zhao02, Jensen05, Samm14, Dirnaichner15}.} These occur owing to changes of the position and the width of the Coulomb resonances, which, in turn, depend on the magnetic configuration (parallel or antiparallel) of the magnetic moments of the contacts, and oscillate with gate and bias voltage~\cite{Dirnaichner15}. The oscillating behavior with gate and bias voltage has also been found in {the strong coupling regime}~\cite{Sahoo05, Man06}, though with a significantly reduced MR amplitude not exceeding {${9}$\%.}

In the present article, we focus on investigating the barrier dependence of the MR effect in CNT spin valves with low-resistive contacts. 
 {The focus is on devices that exhibit metallic conductance.} 
 {Therefore, the devices are built from double-walled (DW) CNTs, since they have a larger probability of being metallic compared to single-walled (SW) CNTs~\cite{Zolyomi08,Koshino15}.} 
We show that one reason for the sample-dependent variation of the MR in CNTs is a difference in strength of the tunnel barrier that forms intrinsically between a contact and a CNT. Depending on the barrier, the MR change can be increased above 10\% even with simple binary alloys used as the contact material  (e.g. CoPd, NiPd, etc.). Identification and optimization of the barrier is therefore of crucial importance, in order to fabricate reliable CNT spin valves with large MR effects. Moreover, we demonstrate a first Hanle measurement on a CNT spin valve that not only corroborates the feasibility of the spin injection in this case, but also shows that such a system competes with state-of-the-art devices based on graphene and Si in terms of the spin lifetime and spin accumulation.

The paper is organized as follows:  in Sec.~\ref{sec:Experimental details}, experimental details of the device fabrication are discussed, as comparability and reproducibility of the devices is crucial for this study. To compare the MR effect of different devices, the measurements have to be performed within the same transport regime as explained in Sec.~\ref{subsec:Temperature and bias}. Next, in Sec.~\ref{subsec:barrier} we present the experimental results on how the MR effect depends on the tunnel barrier strength and introduce the theoretical model to support the interpretation of the data. Finally, spin precession within CNTs is investigated by means of the Hanle effect in Sec.~\ref{subsec:Hanle}.

\section{Experimental details}
\label{sec:Experimental details}
\begin{figure}[t]
  \includegraphics {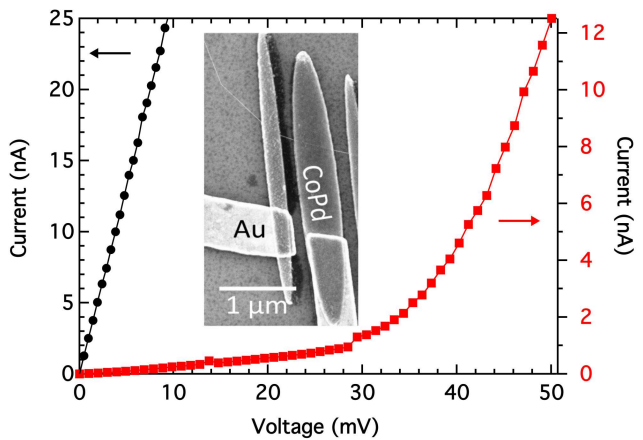}
  \caption{(color online)
  Current-voltage characteristics of a typical device
  under consideration, which exhibits metallic behavior at room temperature (black circles), whereas the  {formation}~of a potential barrier at $4\,$K (red squares) is observed. The inset shows an image of a typical device {taken by a scanning electron microscope}. The position of the tube has been redrawn to enhance visibility.}
  \label{fig1}
\end{figure}

All devices discussed in this work were fabricated on heavily p-doped Si (001) substrates with a $200\,$nm layer thermally oxidized SiO$_2$. Nanotubes were subsequently grown from patterned islands {by} chemical vapor deposition~(CVD) with  an iron-based catalyst~\cite{Kong98}. The contacts were patterned using electron beam lithography. Recently, permalloy and ferromagnet-Pd alloys have shown promising results~\cite{Sahoo05,Kontos10, Preusche09}. At present, we focus our research on CoPd, which has been demonstrated to exhibit a large in-plane magnetization and  {can form low resistive contacts to CNTs~\cite{Morgan12, Morgan13}.} Co and Pd were co-evaporated \emph{via} molecular beam epitaxy to create nanofabricated contacts, while Au was deposited \emph{via} standard metal evaporation to provide coarser leads. A typical resulting lateral spin valve structure is shown in the inset of Fig.~\ref{fig1}. The distance between the CoPd contacts is $150\,$nm and the width of the contacts is $150\,$nm and $400\,$nm respectively, which in turn leads  {to different aspect ratios for the two contacts and in consequence to different coercive fields due to shape anisotropy. A non-trivial temperature dependence of the coercive field has been found and attributed to the interplay of the shape anisotropy and the magnetoelastic effect~\cite{Morgan13}. The influence of this effect on the switching in magnetoresistance measurements is analyzed and discussed below (see Sec.~\ref{subsec:Temperature and bias}).}

The CVD process results in CNTs of mixed chirality, so that both semiconducting as well as metallic tubes are in general expected to be present. Since the aim was to compare similar devices, we chose a growth temperature of $920\,^{\circ}$C that yields mainly double-walled CNTs~\cite{Spudat09}, and thus enhances the probability of fabricating metallic devices. All devices compared in this work showed linear current-voltage characteristics at room temperature, while at $4\,$K a potential barrier becomes apparent, as one can see in Fig.~\ref{fig1}. Note that while this data corresponds to one sample (device~6, cf.~Tab.~\ref{tab1}), it is in fact representative for all devices reported, although the height of the potential barrier varies between different devices.  {This intrinsic tunnel barrier is crucial for measuring a magnetoresistance effect in CNTs, because otherwise the spin polarization would be lost due to the conductivity mismatch~\cite{Rashba00}.}

The interface between the ferromagnetic contact and the CNT is crucial for transport applications. To begin with, we note that the contact interface is quite small which is a consequence of the fact that  {DWCNTs have diameters of a few nanometers (cf. Tab.~\ref{tab1} for the diameters of the CNTs in the devices presented in this paper).} Furthermore, since the injection of charge carriers occurs only at the point where the contact ends~\cite{Tsukagoshi02}, the local magnetic environment has a strong influence on the MR measurements. Importantly, pinning and moving of domain walls in contacts that incorporate several magnetic domains leads to unstable switching behavior of the contacts and a  non-trivial MR signal~\cite{Morgan13a}. For this reason, we optimized the shape of the contacts in such a way that the magnetization is in-plane along the long axis with a single or at most two domains~\cite{Morgan13}.  {The magnetic field in all magnetoresistance measurements presented in this work is applied in direction of the long axis of the contacts unless specified otherwise.}

All measurements discussed here were performed at a temperature of $T = 4\,$K unless specified otherwise. The substrate served as a back gate, and the back gate voltage was kept constant  for all measurements on the same device, as the gate voltage can strongly influence the MR effect~\cite{Sahoo05, Man06}.  In order to be able to compare the magnitude of the MR effect of different devices, measurements presented here were carried out in a regime where the current showed no, or only very weak, dependence on the back gate voltage.

\section{Results and Discussion}
\subsection{Temperature and bias dependence of the magnetoresistance}
\label{subsec:Temperature and bias}

\begin{figure}[bt]
  \includegraphics {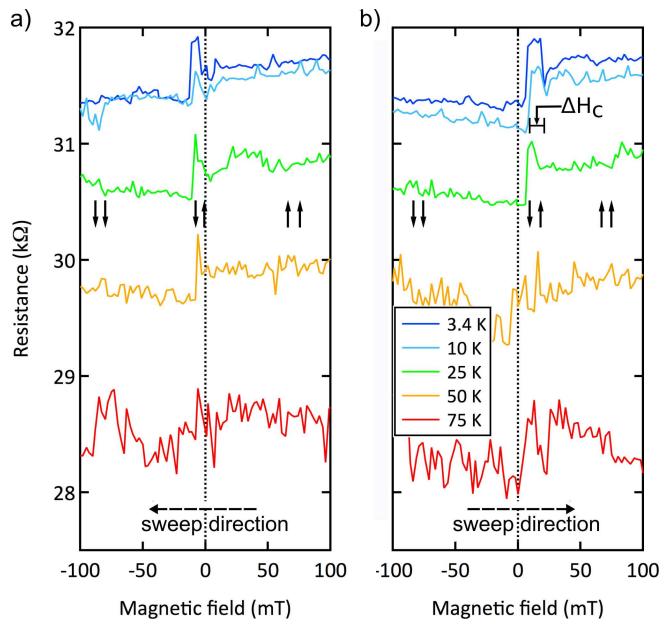}
  \caption{(color online)
  The dependence of local magnetoresistance curves on temperature investigated in a temperature range of $3.4 - 75\,$K  {at $V_\text{bias}=2\,$mV}. The left (right) panel shows results for a field sweep in the negative (positive) direction as indicated by the dashed arrows. The presented data set corresponds to device 2 (cf. Tab.~\ref{tab1}). Solid arrows represent the magnetization direction of the two contacts.}
  \label{fig2}
\end{figure}

Figure~\ref{fig2} shows typical magnetoresistance (MR) measurements taken on a CoPd-contacted CNT in a temperature range from $3.4\,$K up to $75\,$K with a constant bias $V_\text{bias} = 2\,$mV. For the sake of clarity, results for downward and upward field sweeps are plotted separately in~(a) and~(b). The data clearly reveals switching between low and high resistive states for parallel and antiparallel configuration of the magnetization directions of the contacts. As for all devices presented in this work, the MR effect is positive under the given circumstances. The overall resistance is shown to decrease with rising temperature, as expected for electrons tunneling through a barrier~\cite{Lik99}, resulting in the offset observed between the curves. The MR signal decreases in magnitude along with the resistance, all in agreement with previous measurements on Fe-contacted CNTs~\cite{Jensen05}.  {Along with}~the amplitude of the MR switching, the width of the feature is also reduced with increasing temperature from $\Delta H_\text{c} = \pm13\,$mT at $3.4\,$K to $\Delta H_\text{c}\leq4\,$mT at~$50\,$K.  The width of the MR signal is determined by shape anisotropy, as the contacts are fabricated to have different lateral dimensions. The observed temperature-dependent change in width of the MR signal  corresponds to a changing difference in coercive fields of the contacts. This behavior, in turn, correlates well with temperature-dependent SQUID data of CoPd~\cite{Morgan13}, which shows that at $50\,$K,  magnetoelastic anisotropy strongly affects the magnetization of nanocontact arrays, resulting in a lowered coercive field. This effect is different for the two contacts  because of the interplay of shape anisotropy and magnetoelastic anisotropy. The reduced magnitude of the MR signal at elevated temperatures cannot be attributed to a decrease in magnetization of the CoPd contacts since their saturation magnetization is constant within this temperature range~\cite{Morgan13}. Therefore, this is a first indication that the MR effect is due to tunneling and we refer to it as TMR (tunneling magnetoresistance) in the following.

\begin{figure}[bt]
	\includegraphics {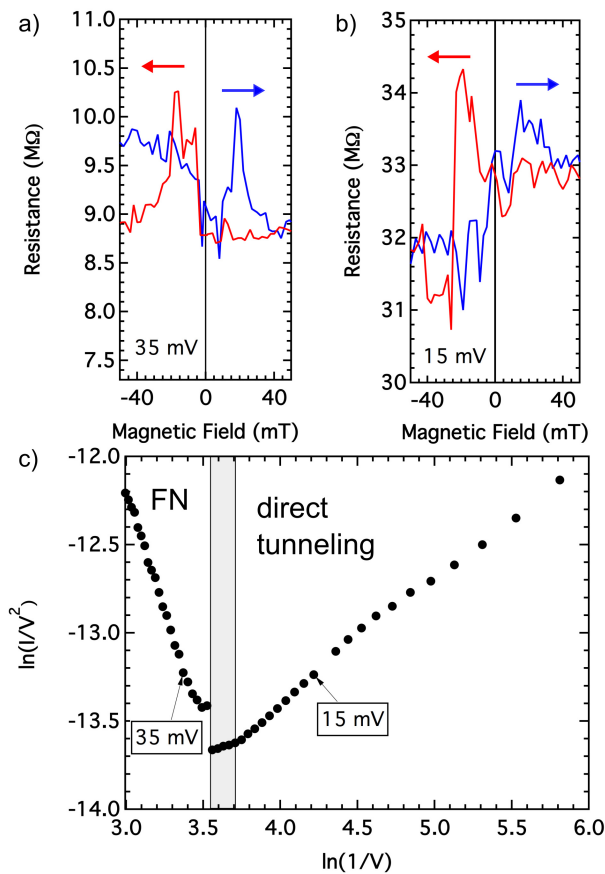}
	\caption{(color online) Magnetoresistance measurements of the sample presented in Fig.~\ref{fig1} obtained at bias voltage: (a) $V_\text{bias}=35\,$mV with TMR = 15\%, and (b) $V_\text{bias}=15\,$mV with TMR = 6\%. Arrows indicate the sweep direction of the magnetic field. (c) The $\ln(I/V^2) - \ln(1/V)$ plot of the current-voltage characteristics at $T = 4\,$K and with $B = 0\,$T reveals that the different sizes of the MR effect can be attributed to different tunneling regimes. The gray shaded region marks the transition regime between direct tunneling (right) and Fowler-Nordheim tunneling (left) that was targeted for the following measurements.
	}
	\label{fig3}
\end{figure}

Figures~\ref{fig3}(a) and~\ref{fig3}(b) show MR measurements of another device at~$4\,$K using different bias voltages $V_\text{bias}$. Though the background of the traces and the details of the TMR switching vary from sweep to sweep and, in particular, between different thermal cycles, the coercive fields and the size of the MR effect observed remain the same.
Introducing the definition of the size of the tunneling magnetoresistance effect
\begin{equation}\label{TMR}
    \text{TMR}
    =
    \frac{G_\textrm{P} - G_\textrm{AP}}{G_\textrm{AP}}
    ,
\end{equation}
with $G_\text{P}$ ($G_\text{AP}$) referring to the conductance in the parallel (antiparallel) magnetic configuration of the contacts, and averaging the backgrounds, we obtain TMR $=15\%$ for the measurement with bias voltage $V_\text{bias} = 35\,$mV [Fig.~\ref{fig3}(a)] and TMR $=6\%$ for the measurement with $V_\text{bias} = 15\,$mV [Fig.~\ref{fig3}(b)]. At first glance, this seems to contradict the general observation that the size of the MR effect in CNTs decreases with increasing bias voltage~\cite{Zhao02, Man06}. Interestingly, the $\ln(I/V^2) - \ln(1/V)$ plot shown in Fig.~\ref{fig3}(c) reveals that the two measurements correspond to different tunneling regimes: the low-bias measurement corresponds to the \emph{direct} tunneling regime, while the high-bias measurement corresponds to the Fowler-Nordheim (FN) tunneling regime~\cite{foot}. 
 {In the direct tunneling regime, the bias voltage applied is smaller than the average barrier height.} 
 {On the other hand, in the}	
 {case of FN tunneling, the bias voltage exceeds the average barrier height.}
 {As a consequence,}
 {the effective barrier width is reduced and the probability of tunneling enhanced.} 
Usually, a $\ln(I/V^2) - 1/V$ plot is used to analyze FN tunneling through an oxide barrier~\cite{Lenzlinger69}. However, the double-logarithmic plot serves to emphasize the transition between the tunneling regimes as the change of slope becomes clearly visible~\cite{Mueller09}. 
 {Note that we do observe oscillations of the conductance with gate voltage at low bias voltage $V\rightarrow 0\,$mV. In contrast to Man \emph{et al.}~\cite{Man06}, the data in this work are} 
 {acquired}
 {at elevated bias voltage, where we do not observe strong oscillations of the conductance. The fact that we work with larger bias voltages is owed to comparably stronger tunnel barriers, which might also explain why in ref. [15] the MR signal vanishes at a bias voltage of 20 mV while our measurements show a large TMR signal at similar bias voltages.}

A decreasing MR effect with increasing bias voltage is typical for conventional magnetic tunnel junctions and organic spin valve structures in the direct tunneling regime~\cite{Moo95, Miy95, Dja05, Zha13}. Spin-selective processes can play a role in the FN tunneling regime either due to band alignment~\cite{Bow06} or trap assisted tunneling~\cite{Qu09} and lead to an increase in TMR~\cite{Mueller09, Nag07}. Before we can understand how these effects might be influenced by the tunnel barrier between the metal contact and a carbon nanotube, we need to find a way to reliably characterize and understand the general influence of this tunnel barrier on the MR effect in CNTs. To exclude an influence of the bias voltage on the signal, we compare the TMR values of different CNT devices measured in the very narrow transition region between these two tunneling regimes, as represented by the gray shaded area in Fig.~\ref{fig3}(c).  This, of course, results in different absolute values of the applied bias voltage, since the conductances of the investigated devices differ.  {The absolute current through all the devices within this regime, however, was the same with $I  = 1\,$nA.}

\subsection{Dependence of magnetoresistance on the strength of the tunnel barrier}
\label{subsec:barrier}

\begin{table*}[tb]
\caption[List of CoPd-contacted devices]{
	Summary of CNT diameters and electronic transport data for the CoPd-contacted CNT devices discussed in this work. In the case of nonohmic current-voltage characteristics at $4\,$K, the resistance listed is determined for the bias voltage used in the transport measurements. Note that the TMR, see Eq.~(\ref{TMR}), is determined at $T = 4\,$K. The error for the TMR value as well as for the barrier strength $Z$,  see Eq.~(\ref{eq1}), is given for the last digit in brackets. It is due to an underestimation of the barrier caused by the temperature dependence of the conductance between room temperature and the temperature where the tunnel barrier forms. Though $1/G(T)$ data exists for all devices, in some cases it was taken at a different bias voltage than the TMR measurement.}
\centering
\renewcommand{\arraystretch}{1.2}
\begin{tabular}{c c c c c c}
\hline\hline
\hspace*{1em}\rule{0cm}{12pt}Device\hspace*{1em}&
\hspace*{1em}CNT diameter (nm)\hspace*{1em}&
\hspace*{1em}1/G($300\,$K$\,\equiv\,$RT) (k$\Omega$)\hspace*{1em}&
\hspace*{1em}1/G($4\,$K)$\equiv {1/G}_\text{P}$ (k$\Omega$)\hspace*{1em}&
\hspace*{1em}TMR~(\%)&$Z$\hspace*{1em}\\[0.5ex]
\hline
\rule{0cm}{12pt}
1 & 5 & 35 & 30 & 0(0.5) & -0.1(+3)\\
2 & 8 & 25 & 32 & 2(0.5) & 0.3(+4)\\
3 & 1.6 & 150 & 210 & 5(0.75) & 0.40(+6)\\
4 & 0.7 & 3700 & 11000 & 6.4(0.5) &\hspace*{1em}1.972(+3)\hspace*{1em}\\
5 & 3.2 & 1000 & 9000 & 12(2)& 8.00(+1)\\
6 & 3 & 350 & 100000 & 15(1) & 24.71(+3)\\[0.5ex]
\hline
\end{tabular}
\label{tab1}
\end{table*}

Table~\ref{tab1} lists the devices used for TMR measurements and it reveals that these devices differ significantly with respect to tube diameter and resistance. First of all, one should note that Pd contacts generally possess a high transmission, which is also known to increase with the diameter of the nanotube even for metallic CNTs~\cite{Kim05}.

Indeed, the  CoPd-contacted devices under investigation exhibit a similar behavior. Devices 1 and 2, characterized by the largest CNT diameters, show a high transmission, though the theoretical  minimal value of the resistance  {$1/G= h/4e^2$}~for a single-wall CNT is not reached. A general trend can be observed that the value of the resistance at room temperature is inversely proportional to the CNT diameter. However, we note that the resistances of devices 5 and 6 appear to be too high compared to device 3.  This deviation can be explained regarding the fact that most of the CNTs used in the present study are double-walled (DW) or even multiwalled, with device 4 being the only exception of a single-walled (SW) CNT. Though devices exhibiting metallic behavior were specifically selected, one should remember that DWCNTs can also be composed of outer semiconducting CNTs, with the inner tubes being metallic. Such a situation would manifest itself in a higher contact resistance and, in particular, as a higher device resistance in the case of our two-terminal device. Moreover, the strong increase in resistance with decreasing temperature supports this argument, as such a device should form a rather strong tunnel barrier. At low temperature, individual CNTs  exhibit intrinsic tunnel barriers of different strength to metal contacts~\cite{Bockrath97}, and this process depends largely on their chirality. Electrons experience a potential barrier at the contact-CNT interface as a result of local hybridization and a local dipolar moment~\cite{Nemec06, Shan04}.

\begin{figure}[bt]
\center \includegraphics {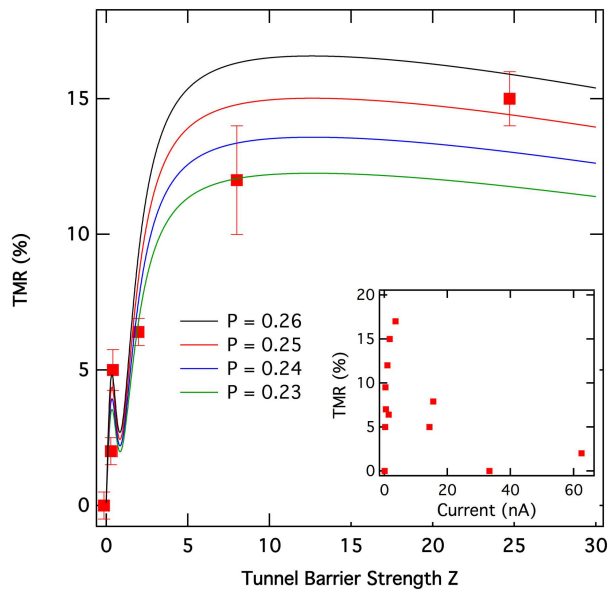}
  \caption{(color online) 
   Evolution of TMR as a function of the tunnel barrier strength $Z$ at $T = 4\,$K. Experimental data ({red squares}) corresponding to different devices (see Tab.~\ref{tab1}) are accompanied by theoretical calculations ({solid lines}) for several values of the magnetic polarization parameter $P$ of injected charge carriers at fixed gate voltage {($\Eg=6\,$meV)}~based on the model explained in the text. Error bars represent deviations in TMR between multiple measurements. The inset shows the magnitude of TMR with respect to current through the {different devices presented}.}
  \label{fig4}
\end{figure}

Figure~\ref{fig4} shows how the {magnitude}~of the TMR {signal}~changes with respect to the strength of the tunnel barrier formed between the CNT and the CoPd contacts. We define the \emph{dimensionless effective barrier strength~$Z$}~of a device by directly comparing the resistance $1/G$ at $4\,$K and at room temperature (RT) as follows  {(see also rightmost column of Table \ref{tab1})}
\begin{equation}\label{eq1}
    Z
    =
   	\frac{1/G(\text{$4\,$K}) - 1/G(\text{RT})}{1/G(\text{RT})}
	={\exp({\frac{E_\text{b}}{k_\text{B}T}}})-1
	\approx
	\frac{E_\text{b}}{k_\text{B}T}
    ,
\end{equation}
with $T= 4\,$K. Here, we assume that the change in resistance {$1/G$}~{with temperature is dominated by tunnel barriers that are overcome at elevated temperature ($T > 4\,$K) by thermal activation of the charge carriers}. This is valid, since all devices investigated exhibit relatively weak temperature dependence during cool-down before the tunnel barrier becomes visible at $T\sim 50\,$K. Physically, the dimensionless barrier strength~$Z$ describes the relation between the energy~$E_\text{b}$ required to overcome the potential barrier and the kinetic (thermal) energy of electrons incident at the contact-CNT interface. We note that when an electron traverses the device it actually encounters two barriers, that is, when it enters and leaves a CNT. These are, however, experimentally indistinguishable and $Z$ includes the overall effect of both tunnel barriers, though essentially $Z$ is determined by the larger one. The normalization with respect to room temperature resistance allows for a direct comparison with the definition of the barrier strength in the model used to interpret our data (see below).

The data indicates that the tunnel barrier strength~$Z$ does indeed influence the magnitude of the TMR as suggested by Slonczewski for conventional TMR~\cite{Slonczewski89}, with devices  characterized by low barriers unable to achieve an effect larger than a few \%. Once the barrier reaches a certain strength, further increase will no longer  affect the TMR. The inset of Fig.~\ref{fig4} shows a plot of TMR vs. current of the same devices that is usually used to analyze the performance of spin valves. Since this type of plot does not account for different tunnel barriers and bias voltage regimes, it is clear that  for our devices a general trend  cannot be expected.

To support our interpretation of the dependence of the TMR on the tunnel barrier strength, we compare our results to model calculations of the TMR for a~CNT, approximated as a ballistic  and  noninteracting one-dimensional (gated) quantum wire~\cite{Kane_Phys.Rev.Lett.78/1997,Egger_Eur.Phys.J.B3/1998,Cottet06} interconnecting two ferromagnetic leads. In essence, such a  model corresponds to an electronic interferometer studied previously both in experiment~\cite{Liang_Nature411/2001,Sahoo05,Man05} and theoretically~\cite{Cottet_Europhys.Lett.74/2006,Cottet06}.
Here, the key element of the model, which has not been addressed so far in full detail, is the effect of the strength of tunnel barriers occurring at the CNT-lead interfaces on spin-dependent transport through the device. As the exact shape of each of the two tunnel barriers is unknown, we describe scattering of tunneling electrons at the CNT-lead interface by means of a repulsive  Dirac-delta potential~$U_0\delta(x)$, the same at both ends of the CNT. This approach has already been proven sufficient to capture some key transport features of the interface for other material systems~\cite{Blonder_Phys.Rev.B25/1982,Qi_Phys.Rev.B58/1998,Grundler_Phys.Rev.Lett.86/2001,Hu_Phys.Rev.Lett.87/2001}.

In general, within the  scattering matrix approach the expression for linear-response conductance of a device at temperature $T$ takes the form~\cite{Datta_book,Blanter_Phys.Rep.336/2000}
    \begin{equation}
    G_{\text{P}/\text{AP}}
    =
    \frac{e^2}{h}
    \sum_{n,\sigma}\int\!\!\text{d}\varepsilon\,
    \mathscr{T}_{n,\sigma}^{\text{P}/\text{AP}}\!(\varepsilon)
    \Bigg(
    \!\!
    -\frac{\partial f(\varepsilon)}{\partial \varepsilon}
    \Bigg)
    ,
    \end{equation}
where $\mathscr{T}_{n,\sigma}^{\text{P}/\text{AP}}\!(\varepsilon)$ denotes the transmission coefficient for an electron of spin $\sigma$ traversing a device through its  {$n$}th channel in the case when the spin moments of the leads are oriented parallel~(P)/antiparallel~(AP), and $f(\varepsilon)=\big[1+\exp\big(\varepsilon/(k_\text{B}T)\big)\big]^{-1}$ is the Fermi-Dirac distribution. Furthermore, we limit our considerations to the case of a
 {CNT,}
which can support only two orbital channels ($n=2$) -- a generic property inherited from graphene which stems from the presence of two inequivalent carbon atoms in a primitive cell~\cite{Kane_Phys.Rev.Lett.78/1997,McEuen_Phys.Rev.Lett.83/1999}. Without losing generality, as long as one assumes two identical and uncoupled channels, a CNT can be approximated as a one-dimensional wire with $\mathscr{T}_{n,\sigma}^c(\varepsilon)=\mathscr{T}_{\sigma}^c(\varepsilon)$ for $c=\text{P},\text{AP}$~\cite{Cottet06}.
Next, the transmission coefficient $\mathscr{T}_{\sigma}^c(\varepsilon)$ is derived in a standard manner by means of the scattering matrix approach~\cite{Datta_book,Veillette_Phys.Rev.B69/2004,Cottet_Europhys.Lett.74/2006}, 
    \begin{equation}
    \mathscr{T}_\sigma^{c}(\varepsilon)
    =
    \frac{
    \mathcal{T}_\sigma\mathcal{T}_{c(\sigma)}
    }{
    \Big|
    1-\sqrt{\big(1-\mathcal{T}_\sigma\big)\big(1-\mathcal{T}_{c(\sigma)}\big)}
    \exp\!\big[i\Phi_\sigma^{c}(\varepsilon)\big]
    \Big|^2
    }
    ,
    \end{equation}
with  $\mathcal{T}_\sigma$ representing the square of the absolute value of the complex transmission amplitude. Note that when writing the equation above we  {assumed}~that both leads and interfaces (represented by the parameter $U_0$) are identical. The action of the magnetic configuration index $c$ on a spin index $\sigma$ is defined as $\text{P}(\sigma)=\sigma$ and $\text{AP}(\sigma)=\overline{\sigma}$, where one should understand the notation $\overline{\sigma}$ as $\overline{\uparrow}\equiv\,\downarrow$ and $\overline{\downarrow}\equiv\,\uparrow$. In the absence of a magnetic field, the phase factor
    $
    \Phi_\sigma^{c}(\varepsilon)
    =
    2\delta(\varepsilon)+\varphi_\sigma+\varphi_{c(\sigma)}
    $
consists of  two terms:  first, $\delta(\varepsilon)=\ell\kw(\varepsilon)$, corresponding to the quantum-mechanical phase acquired by a free electron propagating with a wave vector $\kw(\varepsilon)$ in a wire of length $\ell$, and second, $\varphi_\sigma+\varphi_{c(\sigma)}$, representing the spin-dependent phase shift~\cite{Cottet_Europhys.Lett.74/2006} gained by an electron when it is scattered at the interface back into the wire. If the electrostatic potential energy $|\Eg|$ of the gated wire is small relative to its Fermi energy $\EFw$, one gets $\kw(\varepsilon)=\kFw+(\varepsilon-\Eg)/(\hbar\vFw)$, where energy $\varepsilon$ is defined relative to the Fermi level, $\kFw=8.5\times10^9\,$m$^{-1}$ is a typical value of the Fermi wave vector for electrons in a SWCNT~\cite{Liang_Nature411/2001} and  $\vFw=8\times10^5\,$m/s its Fermi velocity.

Using  standard quantum mechanical methods, and noting that in the low bias voltage limit only electrons  in the vicinity of the Fermi level  of the contacts participate in tunneling,
one obtains~\cite{TheoryPaper}
    \begin{equation}
    \mathcal{T}_\sigma
    =
    \left|
    \frac{
    	2\sqrt{k_\sigma\kw(0)}
    }{
    k_\sigma+\kw(0)+iZ_0\sqrt{\kw(0)\kappa}
    }
    \right|^2
    ,
    \end{equation}
and
    \begin{equation}
    \varphi_\sigma
    =
    \arg\!\left(
    \frac{
    -k_\sigma+\kw(0)-iZ_0\sqrt{\kw(0)\kappa}
    }{
    k_\sigma+\kw(0)+iZ_0\sqrt{\kw(0)\kappa}
    }
    \right)
    .
    \end{equation}   
Here, the spin-dependent wave vector of free electrons in the leads,
	$
	k_{\uparrow(\downarrow)}
	=
	\sqrt{2\me\EF(1\pm 2P)}/\hbar
	$,
is parameterized by the Fermi energy $\EF$ (assumed here $\EF=8.5\,$eV) of the contacts, and the \emph{spin-polarization coefficient} defined as~\cite{Qi_Phys.Rev.B58/1998}
$
	P
	=
	(k_\uparrow-k_\downarrow)/(k_\uparrow+k_\downarrow)
$,
with $P<0.5$.
Furthermore, $\kappa=2\EF/(\hbar\vFw)$ and $Z_0=k_0U_0/\EF$ with $k_0=\sqrt{2\me\EF}/\hbar$ is the \emph{dimensionless barrier strength}, defined as the ratio of  the potential energy of the barrier $k_0U_0$ and the energy of incident electrons from the Fermi level of the contacts, so that it is related 
to the experimental definition of the barrier strength, Eq.~(\ref{eq1}), as $Z=Z_0$.
In consequence, it can be noticed that the only two free parameters of the model are $\Eg$ and $P$.

The fit of the theoretical model to the experimental data points for a specific value of $\Eg$ and different degrees of spin polarization $P$ of the contacts is shown as solid lines in Fig.~\ref{fig4}.  The value of $\Eg$ was chosen so that it matched the experimental finding of only positive values for the MR effect for small $Z$. It is  to some extent  arbitrary, since the MR shows an oscillatory behavior with $\Eg$. 
Moreover, the temperature used for the fit is $T = 3.5\,$K, and the length of the CNT is taken as $\ell = 130\,$nm. Though both these values deviate slightly from those observed in the experiment, they yield better fits. Nevertheless, this seems reasonable, since, first, lithography of various devices is usually not exact, so that it is expected that they are characterized by a somewhat different CNT length, and, second, for a given value of~$\Eg$ and large $Z$ the model is sensitive to the interplay of values of temperature and length~\cite{TheoryPaper}.

Comparing the results of the calculations and the experimental data, we find that the spin polarization of the charge carriers injected from the CoPd contacts is  $24\pm 2\%$. This is rather large compared to CNTs contacted with NiPd, where the spin polarization was estimated to be $\simeq 10$\%~\cite{Man06}. Nevertheless, this is a very reasonable value considering the fact that the polarization of charge carriers injected from pure Co has been found to be between $21\%$ and $35\%$~\cite{Tedrow73} {combined with} the relatively large {polarization of Pd induced by Co~\cite{Shan93}}. 
 {Furthermore, the model is based on a single transport channel while most of the CNTs used in the devices contain two walls and thus likely more transport channels. While the qualitative effect remains unchanged if more channels are included in the model~\cite{TheoryPaper}, this might result in a deviation of the contact polarization. On the other hand, interaction between a CNT and contacts can block channels in a CNT with more than one wall~\cite{Ranjan04} and cancel contributions from other channels.} 
The good agreement of the experimental data with the model further proves the importance of the tunnel barrier strength in optimizing TMR in a CNT-based device.

The fact that the nature and the strength of the barrier {significantly affect} the TMR signal has been observed in ''classical'' metallic tunnel junctions (MTJs) with MgO as the insulating barrier. Generally, the TMR effect increases with increasing MgO thickness~\cite{Yuasa04}. However, pronounced oscillations are observed in MTJs with Heusler alloys~\cite{Marukame10,Wang10} or Fe~\cite{Yuasa04} as the contact material. This is usually attributed to different wave vectors $k$ of the $\Delta_1$ and the $\Delta_5$ band of MgO at the Fermi level that lead to interference effects~\cite{Butler01}. Investigations of graphene spin valves, on the other hand, suggest a monotonic increase of the MR and saturation for strong barriers~\cite{Dlubak12}. 
 {Moreover, the rather small TMR signal previously found in CNT spin valves with permalloy contacts and the small contact polarization deduced using Juli\`ere's model~\cite{Aurich10} might in fact be explained with a weak tunnel barrier in the framework of our model.}

Finally, it should be pointed out that although the proposed model describes the experimental results  relatively well, one should still bear in mind its limitations. The model assumes free (\emph{s}-band) electrons in the contacts, and the CNT-contact interface is approximated by a Dirac-delta potential. In reality, the energy barrier at the interface can be much more complex, with a potential profile depending on many factors, such as the interface roughness or adsorbates. Also, the free-electron model should be applied to transition metals and their alloys with great care, as in principle one should expect a more complicated band structure to be responsible for electron tunneling~\cite{Zhang_Eur.Phys.J.B10/1999}. To account for all these details a model from first principles should be developed~\cite{Mavropoulos_Phys.Rev.B69/2004}, a task which is way beyond the scope of the present paper.

Within these limitations, it is nevertheless obvious that the tunnel barrier has a significant  effect on the TMR  found in CNT spin valves. Furthermore, CoPd-CNT devices form sufficiently low barriers that the addition of tunnel barriers from spin selective insulators becomes feasible. This would also enable one to study the FN tunneling regime more closely, which may can lead to significantly larger TMR values.
	
\subsection{Spin precession within carbon nanotubes}
\label{subsec:Hanle}

As the results discussed above correspond to local measurements, there is the possibility of several spurious effects, such as the Hall effect, anisotropic magnetoresistance (AMR)~\cite{Tombros06}, or other ohmic effects~\cite{Man05} influencing the magnetotransport. One important contribution to the local MR signal can be caused by the magneto-Coulomb effect~\cite{Ono98,Shi03,Sense06}. However, the measurements presented in this work are not taken in the Coulomb blockade regime. For this reason, the influence of the magneto-Coulomb effect should be weak at least. Another effect is the tunneling anisotropic magnetoresistance (TAMR), which can completely mimic the MR effect~\cite{Gould04}. TAMR is an effect solely related to the anisotropic density of states in the ferromagnetic contacts and is usually found in materials with strong spin-orbit coupling that lack inversion symmetry and show AMR, like {lanthanum strontium manganite (LSMO)}~\cite{Gru11}. Though we measure no significant AMR in our CoPd contacts, TAMR might be related to the barrier, as shown for fcc Co(111) contacts on Al$_2$O$_3$~\cite{Wang13}. Since our contacts have a similar structure~\cite{Morgan13}, and the exact nature of the barrier in our devices is unknown, we have to make sure that spin polarized charge carriers are indeed injected into the CNTs.

\begin{figure}[t]
\center \includegraphics {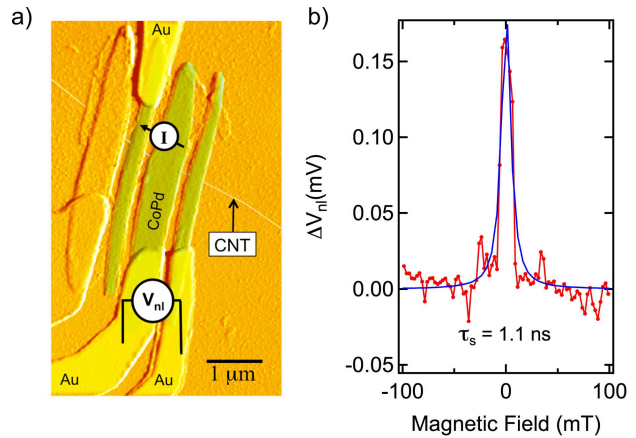}
  \caption{(color online) (a) AFM image of the three-terminal device used in a nonlocal configuration. The CoPd terminals are shaded in green, the gold coarse lines are shaded yellow. The position of the tube has been redrawn for better visibility. Current flow and voltage measurement of the nonlocal signal are schematically indicated. The terminal on the left did not switch its magnetization up to fields as large as 2~T and was left floating. (b) The Hanle measurement (red dots connected by a line that serves merely as a guide to the eyes) was taken at a bias of $I_\text{bias} = 10\,$nA. The fit (blue) with a Lorentz curve yields the spin lifetime of $\tau_\text{s} = 1.1\,$ns.\\
  }
  \label{fig6}
\end{figure}

To prove spin injection into and spin precession within the CNTs, Hanle measurements were performed on a device with a nonlocal three-terminal configuration. Figure~\ref{fig6}(a) presents an {atomic force microscope image (AFM)}~of a device prepared for nonlocal measurements, in which a nonlocal voltage is measured along a separate path from the applied current, using three contacts as indicated. The device displayed a local TMR signal of 15\% and thus, is an example for the strong barrier case as discussed above. For the Hanle measurements, all contacts are magnetized in-plane in the same direction. The sample is then rotated at zero applied field. Subsequently, a magnetic field is applied perpendicular to the plane, which leaves the magnetization of the contacts unaffected due to the shape anisotropy~\cite{Morgan13}. The resulting Hanle signal is presented in Fig.~\ref{fig6}(b).  {Note that this signal was recorded at a bias of  $I_\text{bias} = 10\,$nA and thus outside the Coulomb blockade regime.} The voltage difference $\Delta V_\text{nl}(B)$ measured between the injector and the detector contact at $B=0$ is due to an imbalance of the electrochemical potentials, $\mu^{\uparrow}$ and $\mu^{\downarrow}$, of spin-up and spin-down electrons, respectively.  {It is reduced as the spins dephase while precessing about the perpendicular magnetic field with the Larmor frequency $\omega_\text{L}=g\mu_\text{B} B/\hbar$}. The data is fitted with a Lorentzian peak (blue line), and the magnetic field at the full width at half maximum of the peak is measured to be approximately~$5.2\,$mT. Following Ref.~\cite{Dash09},
    $
    \Delta V_\text{nl}(B)
    =
    \Delta V_\text{nl}(0)/\big[1+(\omega_\text{L} \tau_\text{s})^2\big]
    $
where $\tau_\text{s}$ is the spin lifetime. Taking the \textit{g}-factor \textit{g} = 2 for a CNT in a perpendicularly applied magnetic field~\cite{Cobden98}, we obtain $\tau_\text{s} = 1.1\,$ns.  {This is a lower bound for the spin coherence time $T_2$, since the Hanle effect in combination with spin injection probes the (coherent) dephasing of the conduction electrons~\cite{Zutic04}.} The value of $\Delta V_\text{nl}(0)$ is directly related to the spin accumulation  
    $
    \Delta \mu(0)
    =
    \mu^{\uparrow}-\mu^{\downarrow}
    =
    2e\Delta V_\text{nl}(0)/\mbox{TSP}
    $,
with TSP denoting the tunneling spin polarization of the ferromagnet-barrier interface of  {0.24}~as obtained in the previous section, which yields  {$\Delta \mu(0)\approx1.45\,$meV}. This value is rather large and of similar magnitude to the one found in Si-based spin valves at room temperature~\cite{Dash09}.  The spin lifetime is long compared to the spin lifetime of  $0.1-0.2\,$ns typically found in graphene lateral spin valves on substrates~\cite{Han11,Avsar11}. Only very recently, lifetimes of nanoseconds were reported in high mobility graphene devices sandwiched in hBN~\cite{Guim14} and in suspended graphene devices covered with a layer of BN~\cite{Droegerler14}.

As discussed above, our CNT lateral spin valves compete very well with state-of-the-art silicon or graphene based devices. Because the performance of devices considered in this paper relies on the tunnel barrier that forms  {intrinsically}~between the metal contact and the CNT, the measurements have to be conducted at low temperatures. On the other hand, the spin lifetime in graphene devices is known to exhibit no, or only very little temperature dependence~\cite{Avsar11}. Since we now understand the influence of the tunnel barrier in CNT-based devices, tunnel barrier engineering becomes possible. Spin selective insulators like EuO can be used to increase the temperature for spin injection. The spin lifetime is very likely still influenced by charge traps in the SiO$_2$. Suspending the CNTs or low dose gamma irradiation, which is known to remove traps in the oxide and to improve the performance of CNT-based field-effect transistors~\cite{Sydoruk14}, would be expected to enhance the spin lifetime. The diameter of the CNT used for the Hanle measurements was $1.5\,$nm, giving rise to a spin-orbit coupling strength in the order of $1\,$meV~\cite{Kuemmeth08}. Since spin-orbit coupling is another source of decoherence, using CNTs with larger diameter might increase the spin lifetime as well. 
 {Recently, a coherence time of $60\,$ns was found in a CNT-based double quantum dot coherently coupled to microwave cavity photons~\cite{Viennot15} that indicates the potential of the spin coherence in CNT spin valves. However, electron spin resonance suggests that the electron-electron interactions in a Tomonaga-Luttinger liquid might severely reduce the spin coherence time of one-dimensional metallic devices~\cite{Dora08}. Hanle measurements on improved devices will advance quantitative understanding of spin relaxation properties of CNTs.}

\section{Conclusion}

In this article, we analyzed the TMR effect in lateral spin valve devices based on CNTs. We showed that the strength of the tunnel barrier forming  intrinsically between ferromagnetic contacts made out of CoPd and CNTs has a significant influence on the size of the TMR signal. Approximating a CNT as a ballistic and noninteracting one-dimensional quantum wire, we modeled the experimental data and found that the spin polarization of the injected electrons is about 24\%, and that the TMR effect exhibits a non-monotonic dependence on the barrier strength. Importantly, the usage of CoPd as the contact material allowed for achieving rather high transmission coefficients, with great perspectives for further optimization of the properties of the barrier, for instance by introducing spin-selective insulators. Moreover, Hanle measurements in a nonlocal three-terminal configuration served to confirm the injection of spin-polarized electrons into a CNT. In particular,  spin accumulation and spin lifetime extracted from the Hanle effect turned out to be of similar magnitude as in state-of-the-art spin valve devices based on Si or graphene.  {Our study provides a deeper understanding of the barrier dependence of the TMR in CNTs, facilitating the development of applications of this one-dimensional material in spintronic devices.}

\acknowledgements

The authors thank S.\ Trellenkamp for e-beam writing, T. Jansen for metalization, and  N. Schnitzler,  H.\ Kertz and H.\ Pfeifer for technical assistance with the cryostat setups. We thank D. B\"urgler, G. Schmidt, and P. Mavropoulos for fruitful discussions. The WSxM software was used to analyze all AFM data~\cite{wsxm}. M.M. acknowledges financial support from the Alexander von Humboldt Foundation and the Polish Ministry of Science and Higher Education through a young scientist fellowship (0066/E-336/9/2014). C. Meyer acknowledges financial support from the DFG Research unit FOR912.
%



\begin{thebibliography}{84}%
	\makeatletter
	\providecommand \@ifxundefined [1]{%
		\@ifx{#1\undefined}
	}%
	\providecommand \@ifnum [1]{%
		\ifnum #1\expandafter \@firstoftwo
		\else \expandafter \@secondoftwo
		\fi
	}%
	\providecommand \@ifx [1]{%
		\ifx #1\expandafter \@firstoftwo
		\else \expandafter \@secondoftwo
		\fi
	}%
	\providecommand \natexlab [1]{#1}%
	\providecommand \enquote  [1]{``#1''}%
	\providecommand \bibnamefont  [1]{#1}%
	\providecommand \bibfnamefont [1]{#1}%
	\providecommand \citenamefont [1]{#1}%
	\providecommand \href@noop [0]{\@secondoftwo}%
	\providecommand \href [0]{\begingroup \@sanitize@url \@href}%
	\providecommand \@href[1]{\@@startlink{#1}\@@href}%
	\providecommand \@@href[1]{\endgroup#1\@@endlink}%
	\providecommand \@sanitize@url [0]{\catcode `\\12\catcode `\$12\catcode
		`\&12\catcode `\#12\catcode `\^12\catcode `\_12\catcode `\%12\relax}%
	\providecommand \@@startlink[1]{}%
	\providecommand \@@endlink[0]{}%
	\providecommand \url  [0]{\begingroup\@sanitize@url \@url }%
	\providecommand \@url [1]{\endgroup\@href {#1}{\urlprefix }}%
	\providecommand \urlprefix  [0]{URL }%
	\providecommand \Eprint [0]{\href }%
	\providecommand \doibase [0]{http://dx.doi.org/}%
	\providecommand \selectlanguage [0]{\@gobble}%
	\providecommand \bibinfo  [0]{\@secondoftwo}%
	\providecommand \bibfield  [0]{\@secondoftwo}%
	\providecommand \translation [1]{[#1]}%
	\providecommand \BibitemOpen [0]{}%
	\providecommand \bibitemStop [0]{}%
	\providecommand \bibitemNoStop [0]{.\EOS\space}%
	\providecommand \EOS [0]{\spacefactor3000\relax}%
	\providecommand \BibitemShut  [1]{\csname bibitem#1\endcsname}%
	\let\auto@bib@innerbib\@empty
	\bibitem [{\citenamefont {\ifmmode \check{Z}\else
			\v{Z}\fi{}uti\ifmmode~\acute{c}\else \'{c}\fi{}}\ \emph
		{et~al.}(2004)\citenamefont {\ifmmode \check{Z}\else
			\v{Z}\fi{}uti\ifmmode~\acute{c}\else \'{c}\fi{}}, \citenamefont {Fabian},\
		and\ \citenamefont {Das~Sarma}}]{Zutic04}%
	\BibitemOpen
	\bibfield  {author} {\bibinfo {author} {\bibfnamefont {I.}~\bibnamefont
			{\ifmmode \check{Z}\else \v{Z}\fi{}uti\ifmmode~\acute{c}\else \'{c}\fi{}}},
		\bibinfo {author} {\bibfnamefont {J.}~\bibnamefont {Fabian}}, \ and\ \bibinfo
		{author} {\bibfnamefont {S.}~\bibnamefont {Das~Sarma}},\ }\bibfield  {title}
	{\enquote {\bibinfo {title} {Spintronics: Fundamentals and applications},}\
	}\href@noop {} {\bibfield  {journal} {\bibinfo  {journal} {Rev. Mod. Phys.}\
	}\textbf {\bibinfo {volume} {76}},\ \bibinfo {pages} {323--410} (\bibinfo
	{year} {2004})}\BibitemShut {NoStop}%
\bibitem [{\citenamefont {Han}\ \emph {et~al.}(2014)\citenamefont {Han},
	\citenamefont {Kawakami}, \citenamefont {Gmitra},\ and\ \citenamefont
	{Fabian}}]{Han14}%
\BibitemOpen
\bibfield  {author} {\bibinfo {author} {\bibfnamefont {W.}~\bibnamefont
		{Han}}, \bibinfo {author} {\bibfnamefont {R.~K.}\ \bibnamefont {Kawakami}},
	\bibinfo {author} {\bibfnamefont {M.}~\bibnamefont {Gmitra}}, \ and\ \bibinfo
	{author} {\bibfnamefont {J.}~\bibnamefont {Fabian}},\ }\bibfield  {title}
{\enquote {\bibinfo {title} {Graphene spintronics},}\ }\href@noop {}
{\bibfield  {journal} {\bibinfo  {journal} {Nat. Nanotechnol.}\ }\textbf
	{\bibinfo {volume} {9}},\ \bibinfo {pages} {794--807} (\bibinfo {year}
	{2014})}\BibitemShut {NoStop}%
\bibitem [{\citenamefont {Dr\"ogeler}\ \emph {et~al.}(2014)\citenamefont
	{Dr\"ogeler}, \citenamefont {Volmer}, \citenamefont {Wolter}, \citenamefont
	{Terr\'es}, \citenamefont {Watanabe}, \citenamefont {Taniguchi},
	\citenamefont {G\"untherodt}, \citenamefont {Stampfer},\ and\ \citenamefont
	{Beschoten}}]{Droegerler14}%
\BibitemOpen
\bibfield  {author} {\bibinfo {author} {\bibfnamefont {M.}~\bibnamefont
		{Dr\"ogeler}}, \bibinfo {author} {\bibfnamefont {F.}~\bibnamefont {Volmer}},
	\bibinfo {author} {\bibfnamefont {M.}~\bibnamefont {Wolter}}, \bibinfo
	{author} {\bibfnamefont {B.}~\bibnamefont {Terr\'es}}, \bibinfo {author}
	{\bibfnamefont {K.}~\bibnamefont {Watanabe}}, \bibinfo {author}
	{\bibfnamefont {T.}~\bibnamefont {Taniguchi}}, \bibinfo {author}
	{\bibfnamefont {G.}~\bibnamefont {G\"untherodt}}, \bibinfo {author}
	{\bibfnamefont {C.}~\bibnamefont {Stampfer}}, \ and\ \bibinfo {author}
	{\bibfnamefont {B.}~\bibnamefont {Beschoten}},\ }\bibfield  {title} {\enquote
	{\bibinfo {title} {{Nanosecond spin lifetimes in single-and few-layer
				graphene--hBN heterostructures at room temperature}},}\ }\href@noop {}
{\bibfield  {journal} {\bibinfo  {journal} {Nano Lett.}\ }\textbf {\bibinfo
		{volume} {14}},\ \bibinfo {pages} {6050--6055} (\bibinfo {year}
	{2014})}\BibitemShut {NoStop}%
\bibitem [{\citenamefont {Dlubak}\ \emph {et~al.}(2012)\citenamefont {Dlubak},
	\citenamefont {Martin}, \citenamefont {Deranlot}, \citenamefont {Servet},
	\citenamefont {Xavier}, \citenamefont {Mattana}, \citenamefont {Sprinkle},
	\citenamefont {Berger}, \citenamefont {DeHeer}, \citenamefont {Petroff},
	\citenamefont {Anane}, \citenamefont {Seneor},\ and\ \citenamefont
	{Fert}}]{Dlubak12}%
\BibitemOpen
\bibfield  {author} {\bibinfo {author} {\bibfnamefont {B.}~\bibnamefont
		{Dlubak}}, \bibinfo {author} {\bibfnamefont {M-B.}\ \bibnamefont {Martin}},
	\bibinfo {author} {\bibfnamefont {C.}~\bibnamefont {Deranlot}}, \bibinfo
	{author} {\bibfnamefont {B.}~\bibnamefont {Servet}}, \bibinfo {author}
	{\bibfnamefont {S.}~\bibnamefont {Xavier}}, \bibinfo {author} {\bibfnamefont
		{R.}~\bibnamefont {Mattana}}, \bibinfo {author} {\bibfnamefont
		{M.}~\bibnamefont {Sprinkle}}, \bibinfo {author} {\bibfnamefont
		{C.}~\bibnamefont {Berger}}, \bibinfo {author} {\bibfnamefont {W.~A.}\
		\bibnamefont {DeHeer}}, \bibinfo {author} {\bibfnamefont {F.}~\bibnamefont
		{Petroff}}, \bibinfo {author} {\bibfnamefont {A.}~\bibnamefont {Anane}},
	\bibinfo {author} {\bibfnamefont {P.}~\bibnamefont {Seneor}}, \ and\ \bibinfo
	{author} {\bibfnamefont {A.}~\bibnamefont {Fert}},\ }\bibfield  {title}
{\enquote {\bibinfo {title} {Highly efficient spin transport in epitaxial
			graphene on \protect{SiC}},}\ }\href@noop {} {\bibfield  {journal} {\bibinfo
		{journal} {Nat. Phys.}\ }\textbf {\bibinfo {volume} {8}},\ \bibinfo {pages}
	{557--561} (\bibinfo {year} {2012})}\BibitemShut {NoStop}%
\bibitem [{\citenamefont {Tombros}\ \emph {et~al.}(2007)\citenamefont
	{Tombros}, \citenamefont {Jozsa}, \citenamefont {Popinciuc}, \citenamefont
	{Jonkman},\ and\ \citenamefont {van Wees}}]{Tombros07}%
\BibitemOpen
\bibfield  {author} {\bibinfo {author} {\bibfnamefont {N.}~\bibnamefont
		{Tombros}}, \bibinfo {author} {\bibfnamefont {C.}~\bibnamefont {Jozsa}},
	\bibinfo {author} {\bibfnamefont {M.}~\bibnamefont {Popinciuc}}, \bibinfo
	{author} {\bibfnamefont {H.~T.}\ \bibnamefont {Jonkman}}, \ and\ \bibinfo
	{author} {\bibfnamefont {B.~J.}\ \bibnamefont {van Wees}},\ }\bibfield
{title} {\enquote {\bibinfo {title} {Electronic spin transport and spin
			precession in single graphene layers at room temperature},}\ }\href@noop {}
{\bibfield  {journal} {\bibinfo  {journal} {Nature}\ }\textbf {\bibinfo
		{volume} {448}},\ \bibinfo {pages} {571} (\bibinfo {year}
	{2007})}\BibitemShut {NoStop}%
\bibitem [{\citenamefont {Han}\ \emph {et~al.}(2010)\citenamefont {Han},
	\citenamefont {Pi}, \citenamefont {McCreary}, \citenamefont {Li},
	\citenamefont {Wong}, \citenamefont {Swartz},\ and\ \citenamefont
	{Kawakami}}]{Han10}%
\BibitemOpen
\bibfield  {author} {\bibinfo {author} {\bibfnamefont {W.}~\bibnamefont
		{Han}}, \bibinfo {author} {\bibfnamefont {K.}~\bibnamefont {Pi}}, \bibinfo
	{author} {\bibfnamefont {K.~M.}\ \bibnamefont {McCreary}}, \bibinfo {author}
	{\bibfnamefont {Y.}~\bibnamefont {Li}}, \bibinfo {author} {\bibfnamefont
		{J.~J.~I.}\ \bibnamefont {Wong}}, \bibinfo {author} {\bibfnamefont {A.~G.}\
		\bibnamefont {Swartz}}, \ and\ \bibinfo {author} {\bibfnamefont {R.~K.}\
		\bibnamefont {Kawakami}},\ }\bibfield  {title} {\enquote {\bibinfo {title}
		{Tunneling spin injection into single layer graphene},}\ }\href@noop {}
{\bibfield  {journal} {\bibinfo  {journal} {Phys. Rev. Lett.}\ }\textbf
	{\bibinfo {volume} {105}},\ \bibinfo {pages} {167202} (\bibinfo {year}
	{2010})}\BibitemShut {NoStop}%
\bibitem [{\citenamefont {Kamalakar}\ \emph {et~al.}({2014})\citenamefont
	{Kamalakar}, \citenamefont {Dankert}, \citenamefont {Bergsten}, \citenamefont
	{Ive},\ and\ \citenamefont {Dash}}]{Kamalakar14}%
\BibitemOpen
\bibfield  {author} {\bibinfo {author} {\bibfnamefont {M.~V.}\ \bibnamefont
		{Kamalakar}}, \bibinfo {author} {\bibfnamefont {A.}~\bibnamefont {Dankert}},
	\bibinfo {author} {\bibfnamefont {J.}~\bibnamefont {Bergsten}}, \bibinfo
	{author} {\bibfnamefont {T.}~\bibnamefont {Ive}}, \ and\ \bibinfo {author}
	{\bibfnamefont {S.~P.}\ \bibnamefont {Dash}},\ }\bibfield  {title} {\enquote
	{\bibinfo {title} {{Enhanced Tunnel Spin Injection into Graphene using
				Chemical Vapor Deposited Hexagonal Boron Nitride}},}\ }\href@noop {}
{\bibfield  {journal} {\bibinfo  {journal} {Sci. Rep.}\ }\textbf {\bibinfo
		{volume} {{4}}},\ \bibinfo {pages} {{6146}} (\bibinfo {year}
	{{2014}})}\BibitemShut {NoStop}%
\bibitem [{\citenamefont {T.}\ \emph {et~al.}(1999)\citenamefont {T.},
	\citenamefont {A.},\ and\ \citenamefont {A.}}]{Ago99}%
\BibitemOpen
\bibfield  {author} {\bibinfo {author} {\bibfnamefont {Kazuhito}\
		\bibnamefont {T.}}, \bibinfo {author} {\bibfnamefont {Bruce~W.}\ \bibnamefont
		{A.}}, \ and\ \bibinfo {author} {\bibfnamefont {Hiroki}\ \bibnamefont {A.}},\
}\bibfield  {title} {\enquote {\bibinfo {title} {Coherent transport of
		electron spin in a ferromagnetically contacted carbon nanotube},}\
}\href@noop {} {\bibfield  {journal} {\bibinfo  {journal} {Nature}\ }\textbf
{\bibinfo {volume} {401}},\ \bibinfo {pages} {572} (\bibinfo {year}
{1999})}\BibitemShut {NoStop}%
\bibitem [{\citenamefont {Mann}\ \emph {et~al.}(2003)\citenamefont {Mann},
	\citenamefont {Javey}, \citenamefont {Kong}, \citenamefont {Wang},\ and\
	\citenamefont {Dai}}]{Mann03}%
\BibitemOpen
\bibfield  {author} {\bibinfo {author} {\bibfnamefont {D.}~\bibnamefont
		{Mann}}, \bibinfo {author} {\bibfnamefont {A.}~\bibnamefont {Javey}},
	\bibinfo {author} {\bibfnamefont {J.}~\bibnamefont {Kong}}, \bibinfo {author}
	{\bibfnamefont {Q.}~\bibnamefont {Wang}}, \ and\ \bibinfo {author}
	{\bibfnamefont {H.}~\bibnamefont {Dai}},\ }\bibfield  {title} {\enquote
	{\bibinfo {title} {Ballistic transport in metallic nanotubes with reliable pd
			ohmic contacts},}\ }\href@noop {} {\bibfield  {journal} {\bibinfo  {journal}
		{Nano Lett.}\ }\textbf {\bibinfo {volume} {3}},\ \bibinfo {pages}
	{1541--1544} (\bibinfo {year} {2003})}\BibitemShut {NoStop}%
\bibitem [{\citenamefont {Zhao}\ \emph {et~al.}(2002)\citenamefont {Zhao},
	\citenamefont {M{\"o}nch}, \citenamefont {Vinzelberg}, \citenamefont
	{M{\"u}hl},\ and\ \citenamefont {Schneider}}]{Zhao02}%
\BibitemOpen
\bibfield  {author} {\bibinfo {author} {\bibfnamefont {B.}~\bibnamefont
		{Zhao}}, \bibinfo {author} {\bibfnamefont {I.}~\bibnamefont {M{\"o}nch}},
	\bibinfo {author} {\bibfnamefont {H.}~\bibnamefont {Vinzelberg}}, \bibinfo
	{author} {\bibfnamefont {T.}~\bibnamefont {M{\"u}hl}}, \ and\ \bibinfo
	{author} {\bibfnamefont {C.M.}\ \bibnamefont {Schneider}},\ }\bibfield
{title} {\enquote {\bibinfo {title} {Spin-coherent transport in
			ferromagnetically contacted carbon nanotubes},}\ }\href@noop {} {\bibfield
	{journal} {\bibinfo  {journal} {Appl. Phys. Lett.}\ }\textbf {\bibinfo
		{volume} {80}},\ \bibinfo {pages} {3144--3146} (\bibinfo {year}
	{2002})}\BibitemShut {NoStop}%
\bibitem [{\citenamefont {Jensen}\ \emph {et~al.}(2005)\citenamefont {Jensen},
	\citenamefont {Hauptmann}, \citenamefont {Nyg\aa{}rd},\ and\ \citenamefont
	{Lindelof}}]{Jensen05}%
\BibitemOpen
\bibfield  {author} {\bibinfo {author} {\bibfnamefont {A.}~\bibnamefont
		{Jensen}}, \bibinfo {author} {\bibfnamefont {J.R.}\ \bibnamefont
		{Hauptmann}}, \bibinfo {author} {\bibfnamefont {J.}~\bibnamefont
		{Nyg\aa{}rd}}, \ and\ \bibinfo {author} {\bibfnamefont {P.E.}\ \bibnamefont
		{Lindelof}},\ }\bibfield  {title} {\enquote {\bibinfo {title}
		{Magnetoresistance in ferromagnetically contacted single-wall carbon
			nanotubes},}\ }\href@noop {} {\bibfield  {journal} {\bibinfo  {journal}
		{Phys. Rev. B}\ }\textbf {\bibinfo {volume} {72}},\ \bibinfo {pages} {035419}
	(\bibinfo {year} {2005})}\BibitemShut {NoStop}%
\bibitem [{\citenamefont {Samm}\ \emph {et~al.}(2014)\citenamefont {Samm},
	\citenamefont {Gramich}, \citenamefont {Baumgartner}, \citenamefont {Weiss},\
	and\ \citenamefont {Sch{\"o}nenberger}}]{Samm14}%
\BibitemOpen
\bibfield  {author} {\bibinfo {author} {\bibfnamefont {J.}~\bibnamefont
		{Samm}}, \bibinfo {author} {\bibfnamefont {J.}~\bibnamefont {Gramich}},
	\bibinfo {author} {\bibfnamefont {A.}~\bibnamefont {Baumgartner}}, \bibinfo
	{author} {\bibfnamefont {M.}~\bibnamefont {Weiss}}, \ and\ \bibinfo {author}
	{\bibfnamefont {C.}~\bibnamefont {Sch{\"o}nenberger}},\ }\bibfield  {title}
{\enquote {\bibinfo {title} {Optimized fabrication and characterization of
			carbon nanotube spin valves},}\ }\href@noop {} {\bibfield  {journal}
	{\bibinfo  {journal} {J. Appl. Phys.}\ }\textbf {\bibinfo {volume} {115}},\
	\bibinfo {pages} {174309} (\bibinfo {year} {2014})}\BibitemShut {NoStop}%
\bibitem [{\citenamefont {Dirnaichner}\ \emph {et~al.}(2015)\citenamefont
	{Dirnaichner}, \citenamefont {Grifoni}, \citenamefont {Pr\"ufling},
	\citenamefont {Steininger}, \citenamefont {H\"uttel},\ and\ \citenamefont
	{Strunk}}]{Dirnaichner15}%
\BibitemOpen
\bibfield  {author} {\bibinfo {author} {\bibfnamefont {A.}~\bibnamefont
		{Dirnaichner}}, \bibinfo {author} {\bibfnamefont {M.}~\bibnamefont
		{Grifoni}}, \bibinfo {author} {\bibfnamefont {A.}~\bibnamefont {Pr\"ufling}},
	\bibinfo {author} {\bibfnamefont {D.}~\bibnamefont {Steininger}}, \bibinfo
	{author} {\bibfnamefont {A.K.}\ \bibnamefont {H\"uttel}}, \ and\ \bibinfo
	{author} {\bibfnamefont {C.}~\bibnamefont {Strunk}},\ }\bibfield  {title}
{\enquote {\bibinfo {title} {Transport across a carbon nanotube quantum dot
			contacted with ferromagnetic leads: Experiment and nonperturbative
			modeling},}\ }\href@noop {} {\bibfield  {journal} {\bibinfo  {journal} {Phys.
			Rev. B}\ }\textbf {\bibinfo {volume} {91}},\ \bibinfo {pages} {195402}
	(\bibinfo {year} {2015})}\BibitemShut {NoStop}%
\bibitem [{\citenamefont {Sahoo}\ \emph {et~al.}(2005)\citenamefont {Sahoo},
	\citenamefont {Kontos}, \citenamefont {Furer}, \citenamefont {Hoffmann},
	\citenamefont {Gr{\"a}ber}, \citenamefont {Cottet},\ and\ \citenamefont
	{Sch{\"o}nenberger}}]{Sahoo05}%
\BibitemOpen
\bibfield  {author} {\bibinfo {author} {\bibfnamefont {S.}~\bibnamefont
		{Sahoo}}, \bibinfo {author} {\bibfnamefont {T.}~\bibnamefont {Kontos}},
	\bibinfo {author} {\bibfnamefont {J.}~\bibnamefont {Furer}}, \bibinfo
	{author} {\bibfnamefont {C.}~\bibnamefont {Hoffmann}}, \bibinfo {author}
	{\bibfnamefont {M.}~\bibnamefont {Gr{\"a}ber}}, \bibinfo {author}
	{\bibfnamefont {A.}~\bibnamefont {Cottet}}, \ and\ \bibinfo {author}
	{\bibfnamefont {C.}~\bibnamefont {Sch{\"o}nenberger}},\ }\bibfield  {title}
{\enquote {\bibinfo {title} {Electric field control of spin transport},}\
}\href@noop {} {\bibfield  {journal} {\bibinfo  {journal} {Nat. Phys.}\
}\textbf {\bibinfo {volume} {1}},\ \bibinfo {pages} {99--102} (\bibinfo
{year} {2005})}\BibitemShut {NoStop}%
\bibitem [{\citenamefont {Man}\ \emph {et~al.}(2006)\citenamefont {Man},
	\citenamefont {Wever},\ and\ \citenamefont {Morpurgo}}]{Man06}%
\BibitemOpen
\bibfield  {author} {\bibinfo {author} {\bibfnamefont {H.T.}\ \bibnamefont
		{Man}}, \bibinfo {author} {\bibfnamefont {I.J.W.}\ \bibnamefont {Wever}}, \
	and\ \bibinfo {author} {\bibfnamefont {A.F.}\ \bibnamefont {Morpurgo}},\
}\bibfield  {title} {\enquote {\bibinfo {title} {Spin-dependent quantum
		interference in single-wall carbon nanotubes with ferromagnetic contacts},}\
}\href@noop {} {\bibfield  {journal} {\bibinfo  {journal} {Phys. Rev. B}\
}\textbf {\bibinfo {volume} {73}},\ \bibinfo {pages} {241401} (\bibinfo
{year} {2006})}\BibitemShut {NoStop}%
\bibitem [{\citenamefont {Z\'olyomi}\ \emph {et~al.}(2008)\citenamefont
	{Z\'olyomi}, \citenamefont {Koltai}, \citenamefont {Ruszny\'ak},
	\citenamefont {K\"urti}, \citenamefont {Gali}, \citenamefont {Simon},
	\citenamefont {Kuzmany}, \citenamefont {Szabados},\ and\ \citenamefont
	{Surj\'an}}]{Zolyomi08}%
\BibitemOpen
\bibfield  {author} {\bibinfo {author} {\bibfnamefont {V.}~\bibnamefont
		{Z\'olyomi}}, \bibinfo {author} {\bibfnamefont {J.}~\bibnamefont {Koltai}},
	\bibinfo {author} {\bibfnamefont {\'A.}\ \bibnamefont {Ruszny\'ak}}, \bibinfo
	{author} {\bibfnamefont {J.}~\bibnamefont {K\"urti}}, \bibinfo {author}
	{\bibfnamefont {\'A.}\ \bibnamefont {Gali}}, \bibinfo {author} {\bibfnamefont
		{F.}~\bibnamefont {Simon}}, \bibinfo {author} {\bibfnamefont
		{H.}~\bibnamefont {Kuzmany}}, \bibinfo {author} {\bibfnamefont {\'A.}\
		\bibnamefont {Szabados}}, \ and\ \bibinfo {author} {\bibfnamefont {P.~R.}\
		\bibnamefont {Surj\'an}},\ }\bibfield  {title} {\enquote {\bibinfo {title}
		{Intershell interaction in double walled carbon nanotubes: Charge transfer
			and orbital mixing},}\ }\href@noop {} {\bibfield  {journal} {\bibinfo
		{journal} {Phys. Rev. B}\ }\textbf {\bibinfo {volume} {77}},\ \bibinfo
	{pages} {245403} (\bibinfo {year} {2008})}\BibitemShut {NoStop}%
\bibitem [{\citenamefont {Koshino}\ \emph {et~al.}(2015)\citenamefont
	{Koshino}, \citenamefont {Moon},\ and\ \citenamefont {Son}}]{Koshino15}%
\BibitemOpen
\bibfield  {author} {\bibinfo {author} {\bibfnamefont {M.}~\bibnamefont
		{Koshino}}, \bibinfo {author} {\bibfnamefont {P.}~\bibnamefont {Moon}}, \
	and\ \bibinfo {author} {\bibfnamefont {Y.-W.}\ \bibnamefont {Son}},\
}\bibfield  {title} {\enquote {\bibinfo {title} {Incommensurate double-walled
		carbon nanotubes as one-dimensional moir\'e crystals},}\ }\href@noop {}
{\bibfield  {journal} {\bibinfo  {journal} {Phys. Rev. B}\ }\textbf {\bibinfo
		{volume} {91}},\ \bibinfo {pages} {035405} (\bibinfo {year}
	{2015})}\BibitemShut {NoStop}%
\bibitem [{\citenamefont {Kong}\ \emph {et~al.}(1998)\citenamefont {Kong},
	\citenamefont {Soh}, \citenamefont {Cassell}, \citenamefont {Quate},\ and\
	\citenamefont {Dai}}]{Kong98}%
\BibitemOpen
\bibfield  {author} {\bibinfo {author} {\bibfnamefont {J.}~\bibnamefont
		{Kong}}, \bibinfo {author} {\bibfnamefont {H.~T.}\ \bibnamefont {Soh}},
	\bibinfo {author} {\bibfnamefont {A.~M.}\ \bibnamefont {Cassell}}, \bibinfo
	{author} {\bibfnamefont {C.~F.}\ \bibnamefont {Quate}}, \ and\ \bibinfo
	{author} {\bibfnamefont {H.~J.}\ \bibnamefont {Dai}},\ }\bibfield  {title}
{\enquote {\bibinfo {title} {Synthesis of individual single-walled carbon
			nanotubes on patterned silicon wafers},}\ }\href@noop {} {\bibfield
	{journal} {\bibinfo  {journal} {Nature}\ }\textbf {\bibinfo {volume} {395}},\
	\bibinfo {pages} {878--881} (\bibinfo {year} {1998})}\BibitemShut {NoStop}%
\bibitem [{\citenamefont {Feuillet-Palma}\ \emph {et~al.}(2010)\citenamefont
	{Feuillet-Palma}, \citenamefont {Delattre}, \citenamefont {Morfin},
	\citenamefont {Berroir}, \citenamefont {F\`eve}, \citenamefont {Glattli},
	\citenamefont {Pla\ifmmode~\mbox{\c{c}}\else \c{c}\fi{}ais}, \citenamefont
	{Cottet},\ and\ \citenamefont {Kontos}}]{Kontos10}%
\BibitemOpen
\bibfield  {author} {\bibinfo {author} {\bibfnamefont {C.}~\bibnamefont
		{Feuillet-Palma}}, \bibinfo {author} {\bibfnamefont {T.}~\bibnamefont
		{Delattre}}, \bibinfo {author} {\bibfnamefont {P.}~\bibnamefont {Morfin}},
	\bibinfo {author} {\bibfnamefont {J.-M.}\ \bibnamefont {Berroir}}, \bibinfo
	{author} {\bibfnamefont {G.}~\bibnamefont {F\`eve}}, \bibinfo {author}
	{\bibfnamefont {D.~C.}\ \bibnamefont {Glattli}}, \bibinfo {author}
	{\bibfnamefont {B.}~\bibnamefont {Pla\ifmmode~\mbox{\c{c}}\else
			\c{c}\fi{}ais}}, \bibinfo {author} {\bibfnamefont {A.}~\bibnamefont
		{Cottet}}, \ and\ \bibinfo {author} {\bibfnamefont {T.}~\bibnamefont
		{Kontos}},\ }\bibfield  {title} {\enquote {\bibinfo {title} {Conserved spin
			and orbital phase along carbon nanotubes connected with multiple
			ferromagnetic contacts},}\ }\href@noop {} {\bibfield  {journal} {\bibinfo
		{journal} {Phys. Rev. B}\ }\textbf {\bibinfo {volume} {81}},\ \bibinfo
	{pages} {115414} (\bibinfo {year} {2010})}\BibitemShut {NoStop}%
\bibitem [{\citenamefont {Preusche}\ \emph {et~al.}(2009)\citenamefont
	{Preusche}, \citenamefont {Schmidmeier}, \citenamefont {Pallecchi},
	\citenamefont {Dietrich}, \citenamefont {H\"{u}ttel}, \citenamefont {Zweck},\
	and\ \citenamefont {Strunk}}]{Preusche09}%
\BibitemOpen
\bibfield  {author} {\bibinfo {author} {\bibfnamefont {D.}~\bibnamefont
		{Preusche}}, \bibinfo {author} {\bibfnamefont {S.}~\bibnamefont
		{Schmidmeier}}, \bibinfo {author} {\bibfnamefont {E.}~\bibnamefont
		{Pallecchi}}, \bibinfo {author} {\bibfnamefont {C.}~\bibnamefont {Dietrich}},
	\bibinfo {author} {\bibfnamefont {A.K.}\ \bibnamefont {H\"{u}ttel}}, \bibinfo
	{author} {\bibfnamefont {J.}~\bibnamefont {Zweck}}, \ and\ \bibinfo {author}
	{\bibfnamefont {C.}~\bibnamefont {Strunk}},\ }\bibfield  {title} {\enquote
	{\bibinfo {title} {Characterization of ferromagnetic contacts to carbon
			nanotubes},}\ }\href@noop {} {\bibfield  {journal} {\bibinfo  {journal} {J.
			Appl. Phys.}\ }\textbf {\bibinfo {volume} {106}},\ \bibinfo {pages} {084314}
	(\bibinfo {year} {2009})}\BibitemShut {NoStop}%
\bibitem [{\citenamefont {Morgan}\ \emph {et~al.}(2012)\citenamefont {Morgan},
	\citenamefont {Schneider},\ and\ \citenamefont {Meyer}}]{Morgan12}%
\BibitemOpen
\bibfield  {author} {\bibinfo {author} {\bibfnamefont {C.}~\bibnamefont
		{Morgan}}, \bibinfo {author} {\bibfnamefont {C.M.}\ \bibnamefont
		{Schneider}}, \ and\ \bibinfo {author} {\bibfnamefont {C.}~\bibnamefont
		{Meyer}},\ }\bibfield  {title} {\enquote {\bibinfo {title} {{Permalloy and
				Co$_{50}$Pd$_{50}$ as ferromagnetic contacts for magnetoresistance
				measurements in carbon nanotube-based transport structures}},}\ }\href@noop
{} {\bibfield  {journal} {\bibinfo  {journal} {J. Appl. Phys.}\ }\textbf
	{\bibinfo {volume} {111}},\ \bibinfo {pages} {07B309} (\bibinfo {year}
	{2012})}\BibitemShut {NoStop}%
\bibitem [{\citenamefont {Morgan}\ \emph
	{et~al.}(2013{\natexlab{a}})\citenamefont {Morgan}, \citenamefont
	{Schmalbuch}, \citenamefont {Garc\'ia-S\'anchez}, \citenamefont {Schneider},\
	and\ \citenamefont {Meyer}}]{Morgan13}%
\BibitemOpen
\bibfield  {author} {\bibinfo {author} {\bibfnamefont {C.}~\bibnamefont
		{Morgan}}, \bibinfo {author} {\bibfnamefont {K.}~\bibnamefont {Schmalbuch}},
	\bibinfo {author} {\bibfnamefont {F}~\bibnamefont {Garc\'ia-S\'anchez}},
	\bibinfo {author} {\bibfnamefont {C.M.}\ \bibnamefont {Schneider}}, \ and\
	\bibinfo {author} {\bibfnamefont {C.}~\bibnamefont {Meyer}},\ }\bibfield
{title} {\enquote {\bibinfo {title} {{Structure and magnetization in CoPd
				thin films and nanocontacts}},}\ }\href@noop {} {\bibfield  {journal}
	{\bibinfo  {journal} {J. Magn. Magn. Mater.}\ }\textbf {\bibinfo {volume}
		{325}},\ \bibinfo {pages} {112--116} (\bibinfo {year}
	{2013}{\natexlab{a}})}\BibitemShut {NoStop}%
\bibitem [{\citenamefont {Spudat}\ \emph {et~al.}(2009)\citenamefont {Spudat},
	\citenamefont {Meyer}, \citenamefont {Goss},\ and\ \citenamefont
	{Schneider}}]{Spudat09}%
\BibitemOpen
\bibfield  {author} {\bibinfo {author} {\bibfnamefont {C.}~\bibnamefont
		{Spudat}}, \bibinfo {author} {\bibfnamefont {C.}~\bibnamefont {Meyer}},
	\bibinfo {author} {\bibfnamefont {K.}~\bibnamefont {Goss}}, \ and\ \bibinfo
	{author} {\bibfnamefont {C.M.}\ \bibnamefont {Schneider}},\ }\bibfield
{title} {\enquote {\bibinfo {title} {Peapod synthesis depending on the number
			of nanotube sidewalls},}\ }\href@noop {} {\bibfield  {journal} {\bibinfo
		{journal} {Phys. Status Solidi B}\ }\textbf {\bibinfo {volume} {246}},\
	\bibinfo {pages} {2498--2501} (\bibinfo {year} {2009})}\BibitemShut {NoStop}%
\bibitem [{\citenamefont {Rashba}(2000)}]{Rashba00}%
\BibitemOpen
\bibfield  {author} {\bibinfo {author} {\bibfnamefont {E.~I.}\ \bibnamefont
		{Rashba}},\ }\bibfield  {title} {\enquote {\bibinfo {title} {Theory of
			electrical spin injection: Tunnel contacts as a solution of the conductivity
			mismatch problem},}\ }\href@noop {} {\bibfield  {journal} {\bibinfo
		{journal} {Phys. Rev. B}\ }\textbf {\bibinfo {volume} {62}},\ \bibinfo
	{pages} {R16267--R16270} (\bibinfo {year} {2000})}\BibitemShut {NoStop}%
\bibitem [{\citenamefont {Tsukagoshi}\ \emph {et~al.}(2002)\citenamefont
	{Tsukagoshi}, \citenamefont {Yoneya}, \citenamefont {Uryu}, \citenamefont
	{Aoyagi}, \citenamefont {Kanda}, \citenamefont {Ootuka},\ and\ \citenamefont
	{Alphenaar}}]{Tsukagoshi02}%
\BibitemOpen
\bibfield  {author} {\bibinfo {author} {\bibfnamefont {K.}~\bibnamefont
		{Tsukagoshi}}, \bibinfo {author} {\bibfnamefont {N.}~\bibnamefont {Yoneya}},
	\bibinfo {author} {\bibfnamefont {S.}~\bibnamefont {Uryu}}, \bibinfo {author}
	{\bibfnamefont {Y.}~\bibnamefont {Aoyagi}}, \bibinfo {author} {\bibfnamefont
		{A.}~\bibnamefont {Kanda}}, \bibinfo {author} {\bibfnamefont
		{Y.}~\bibnamefont {Ootuka}}, \ and\ \bibinfo {author} {\bibfnamefont {B.W.}\
		\bibnamefont {Alphenaar}},\ }\bibfield  {title} {\enquote {\bibinfo {title}
		{Carbon nanotube devices for nanoelectronics},}\ }\href@noop {} {\bibfield
	{journal} {\bibinfo  {journal} {Physica B}\ }\textbf {\bibinfo {volume}
		{323}},\ \bibinfo {pages} {107--114} (\bibinfo {year} {2002})}\BibitemShut
{NoStop}%
\bibitem [{\citenamefont {Morgan}\ \emph
	{et~al.}(2013{\natexlab{b}})\citenamefont {Morgan}, \citenamefont {Metten},
	\citenamefont {Schneider},\ and\ \citenamefont {Meyer}}]{Morgan13a}%
\BibitemOpen
\bibfield  {author} {\bibinfo {author} {\bibfnamefont {C.}~\bibnamefont
		{Morgan}}, \bibinfo {author} {\bibfnamefont {D.}~\bibnamefont {Metten}},
	\bibinfo {author} {\bibfnamefont {C.~M.}\ \bibnamefont {Schneider}}, \ and\
	\bibinfo {author} {\bibfnamefont {C.}~\bibnamefont {Meyer}},\ }\bibfield
{title} {\enquote {\bibinfo {title} {Effect of contact geometry on
			magnetoresistance in copd-contacted carbon nanotubes},}\ }\href@noop {}
{\bibfield  {journal} {\bibinfo  {journal} {Phys. Status Solidi B}\ }\textbf
	{\bibinfo {volume} {250}},\ \bibinfo {pages} {2622--2626} (\bibinfo {year}
	{2013}{\natexlab{b}})}\BibitemShut {NoStop}%
\bibitem [{\citenamefont {Likharev}(1999)}]{Lik99}%
\BibitemOpen
\bibfield  {author} {\bibinfo {author} {\bibfnamefont {K.K.}\ \bibnamefont
		{Likharev}},\ }\bibfield  {title} {\enquote {\bibinfo {title}
		{Single-electron devices and their applications},}\ }\href@noop {} {\bibfield
	{journal} {\bibinfo  {journal} {Proc. IEEE}\ }\textbf {\bibinfo {volume}
		{87}},\ \bibinfo {pages} {606--632} (\bibinfo {year} {1999})}\BibitemShut
{NoStop}%
\bibitem [{foo()}]{foot}%
\BibitemOpen
\href@noop {} {}\bibinfo {note} {The``jump'' that appears between the
	transition regime and the FN regime is caused by a internal switching of
	measurement ranges in the Keithley 2636B at a current of $I = 1\,$nA. It does
	not affect the qualitative shape of the curve (cf.
	Fig.~\ref{fig1}).}\BibitemShut {Stop}%
\bibitem [{\citenamefont {Lenzlinger}\ and\ \citenamefont
	{Snow}(1969)}]{Lenzlinger69}%
\BibitemOpen
\bibfield  {author} {\bibinfo {author} {\bibfnamefont {M.}~\bibnamefont
		{Lenzlinger}}\ and\ \bibinfo {author} {\bibfnamefont {E.~H.}\ \bibnamefont
		{Snow}},\ }\bibfield  {title} {\enquote {\bibinfo {title} {{Fowler-Nordheim
				tunneling into thermally grown SiO$_2$}},}\ }\href@noop {} {\bibfield
	{journal} {\bibinfo  {journal} {J. Appl. Phys.}\ }\textbf {\bibinfo {volume}
		{40}},\ \bibinfo {pages} {278--283} (\bibinfo {year} {1969})}\BibitemShut
{NoStop}%
\bibitem [{\citenamefont {M\"uller}\ \emph {et~al.}(2009)\citenamefont
	{M\"uller}, \citenamefont {Miao},\ and\ \citenamefont {Moodera}}]{Mueller09}%
\BibitemOpen
\bibfield  {author} {\bibinfo {author} {\bibfnamefont {M.}~\bibnamefont
		{M\"uller}}, \bibinfo {author} {\bibfnamefont {G.-X.}\ \bibnamefont {Miao}},
	\ and\ \bibinfo {author} {\bibfnamefont {J.S.}\ \bibnamefont {Moodera}},\
}\bibfield  {title} {\enquote {\bibinfo {title} {Exchange splitting and
		bias-dependent transport in \protect{EuO} spin filter tunnel barriers},}\
}\href@noop {} {\bibfield  {journal} {\bibinfo  {journal} {Europhys. Lett.}\
}\textbf {\bibinfo {volume} {88}},\ \bibinfo {pages} {47006} (\bibinfo {year}
{2009})}\BibitemShut {NoStop}%
\bibitem [{\citenamefont {Moodera}\ \emph {et~al.}(1995)\citenamefont
	{Moodera}, \citenamefont {Kinder}, \citenamefont {Wong},\ and\ \citenamefont
	{Meservey}}]{Moo95}%
\BibitemOpen
\bibfield  {author} {\bibinfo {author} {\bibfnamefont {J.S.}\ \bibnamefont
		{Moodera}}, \bibinfo {author} {\bibfnamefont {L.R.}\ \bibnamefont {Kinder}},
	\bibinfo {author} {\bibfnamefont {T.M.}\ \bibnamefont {Wong}}, \ and\
	\bibinfo {author} {\bibfnamefont {R.}~\bibnamefont {Meservey}},\ }\bibfield
{title} {\enquote {\bibinfo {title} {Large magnetoresistance at room
			temperature in ferromagnetic thin film tunnel junctions},}\ }\href@noop {}
{\bibfield  {journal} {\bibinfo  {journal} {Phys. Rev. Lett.}\ }\textbf
	{\bibinfo {volume} {74}},\ \bibinfo {pages} {3273--3276} (\bibinfo {year}
	{1995})}\BibitemShut {NoStop}%
\bibitem [{\citenamefont {Miyazaki}\ and\ \citenamefont
	{Tezuka}(1995)}]{Miy95}%
\BibitemOpen
\bibfield  {author} {\bibinfo {author} {\bibfnamefont {T.}~\bibnamefont
		{Miyazaki}}\ and\ \bibinfo {author} {\bibfnamefont {N.}~\bibnamefont
		{Tezuka}},\ }\bibfield  {title} {\enquote {\bibinfo {title} {{Giant magnetic
				tunneling effect in Fe/Al$_2$O$_3$/Fe junction }},}\ }\href@noop {}
{\bibfield  {journal} {\bibinfo  {journal} {J. Magn. Magn. Mater.}\ }\textbf
	{\bibinfo {volume} {139}},\ \bibinfo {pages} {L231 -- L234} (\bibinfo {year}
	{1995})}\BibitemShut {NoStop}%
\bibitem [{\citenamefont {Djayaprawira}\ \emph {et~al.}(2005)\citenamefont
	{Djayaprawira}, \citenamefont {Tsunekawa}, \citenamefont {Nagai},
	\citenamefont {Maehara}, \citenamefont {Yamagata}, \citenamefont {Watanabe},
	\citenamefont {Yuasa}, \citenamefont {Suzuki},\ and\ \citenamefont
	{Ando}}]{Dja05}%
\BibitemOpen
\bibfield  {author} {\bibinfo {author} {\bibfnamefont {D.D.}\ \bibnamefont
		{Djayaprawira}}, \bibinfo {author} {\bibfnamefont {K.}~\bibnamefont
		{Tsunekawa}}, \bibinfo {author} {\bibfnamefont {M.}~\bibnamefont {Nagai}},
	\bibinfo {author} {\bibfnamefont {H.}~\bibnamefont {Maehara}}, \bibinfo
	{author} {\bibfnamefont {S.}~\bibnamefont {Yamagata}}, \bibinfo {author}
	{\bibfnamefont {N.}~\bibnamefont {Watanabe}}, \bibinfo {author}
	{\bibfnamefont {S.}~\bibnamefont {Yuasa}}, \bibinfo {author} {\bibfnamefont
		{Y.}~\bibnamefont {Suzuki}}, \ and\ \bibinfo {author} {\bibfnamefont
		{K.}~\bibnamefont {Ando}},\ }\bibfield  {title} {\enquote {\bibinfo {title}
		{{230\% room-temperature magnetoresistance in CoFeB/MgO/CoFeB magnetic tunnel
				junctions}},}\ }\href@noop {} {\bibfield  {journal} {\bibinfo  {journal}
		{Appl. Phys. Lett.}\ }\textbf {\bibinfo {volume} {86}},\ \bibinfo {pages}
	{092502} (\bibinfo {year} {2005})}\BibitemShut {NoStop}%
\bibitem [{\citenamefont {Zhang}\ \emph {et~al.}(2013)\citenamefont {Zhang},
	\citenamefont {Mizukami}, \citenamefont {Kubota}, \citenamefont {Ma},
	\citenamefont {Oogane}, \citenamefont {Naganuma}, \citenamefont {Ando},\ and\
	\citenamefont {Miyazaki}}]{Zha13}%
\BibitemOpen
\bibfield  {author} {\bibinfo {author} {\bibfnamefont {X.}~\bibnamefont
		{Zhang}}, \bibinfo {author} {\bibfnamefont {S.}~\bibnamefont {Mizukami}},
	\bibinfo {author} {\bibfnamefont {T.}~\bibnamefont {Kubota}}, \bibinfo
	{author} {\bibfnamefont {Q.}~\bibnamefont {Ma}}, \bibinfo {author}
	{\bibfnamefont {M.}~\bibnamefont {Oogane}}, \bibinfo {author} {\bibfnamefont
		{H.}~\bibnamefont {Naganuma}}, \bibinfo {author} {\bibfnamefont
		{Y.}~\bibnamefont {Ando}}, \ and\ \bibinfo {author} {\bibfnamefont
		{T.}~\bibnamefont {Miyazaki}},\ }\bibfield  {title} {\enquote {\bibinfo
		{title} {Observation of a large spin-dependent transport length in organic
			spin valves at room temperature},}\ }\href@noop {} {\bibfield  {journal}
	{\bibinfo  {journal} {Nat. Commun.}\ }\textbf {\bibinfo {volume} {4}},\
	\bibinfo {pages} {1392} (\bibinfo {year} {2013})}\BibitemShut {NoStop}%
\bibitem [{\citenamefont {Bowen}\ \emph {et~al.}(2006)\citenamefont {Bowen},
	\citenamefont {Barth\'el\'emy}, \citenamefont {Bellini}, \citenamefont
	{Bibes}, \citenamefont {Seneor}, \citenamefont {Jacquet}, \citenamefont
	{Contour},\ and\ \citenamefont {Dederichs}}]{Bow06}%
\BibitemOpen
\bibfield  {author} {\bibinfo {author} {\bibfnamefont {M.}~\bibnamefont
		{Bowen}}, \bibinfo {author} {\bibfnamefont {A.}~\bibnamefont
		{Barth\'el\'emy}}, \bibinfo {author} {\bibfnamefont {V.}~\bibnamefont
		{Bellini}}, \bibinfo {author} {\bibfnamefont {M.}~\bibnamefont {Bibes}},
	\bibinfo {author} {\bibfnamefont {P.}~\bibnamefont {Seneor}}, \bibinfo
	{author} {\bibfnamefont {E.}~\bibnamefont {Jacquet}}, \bibinfo {author}
	{\bibfnamefont {J.-P.}\ \bibnamefont {Contour}}, \ and\ \bibinfo {author}
	{\bibfnamefont {P.H.}\ \bibnamefont {Dederichs}},\ }\bibfield  {title}
{\enquote {\bibinfo {title} {{Observation of Fowler-Nordheim hole tunneling
				across an electron tunnel junction due to total symmetry filtering}},}\
}\href@noop {} {\bibfield  {journal} {\bibinfo  {journal} {Phys. Rev. B}\
}\textbf {\bibinfo {volume} {73}},\ \bibinfo {pages} {140408} (\bibinfo
{year} {2006})}\BibitemShut {NoStop}%
\bibitem [{\citenamefont {Qu}\ \emph {et~al.}({2009})\citenamefont {Qu},
	\citenamefont {Li}, \citenamefont {Zhao}, \citenamefont {Mei}, \citenamefont
	{Liu}, \citenamefont {Tian}, \citenamefont {Shi}, \citenamefont {Guo},
	\citenamefont {Li}, \citenamefont {Zheng},\ and\ \citenamefont {Li}}]{Qu09}%
\BibitemOpen
\bibfield  {author} {\bibinfo {author} {\bibfnamefont {T.L.}\ \bibnamefont
		{Qu}}, \bibinfo {author} {\bibfnamefont {J.}~\bibnamefont {Li}}, \bibinfo
	{author} {\bibfnamefont {Y.G.}\ \bibnamefont {Zhao}}, \bibinfo {author}
	{\bibfnamefont {J.W.}\ \bibnamefont {Mei}}, \bibinfo {author} {\bibfnamefont
		{X.}~\bibnamefont {Liu}}, \bibinfo {author} {\bibfnamefont {H.F.}\
		\bibnamefont {Tian}}, \bibinfo {author} {\bibfnamefont {J.P.}\ \bibnamefont
		{Shi}}, \bibinfo {author} {\bibfnamefont {S.M.}\ \bibnamefont {Guo}},
	\bibinfo {author} {\bibfnamefont {J.}~\bibnamefont {Li}}, \bibinfo {author}
	{\bibfnamefont {D.N.}\ \bibnamefont {Zheng}}, \ and\ \bibinfo {author}
	{\bibfnamefont {J.Q.}\ \bibnamefont {Li}},\ }\bibfield  {title} {\enquote
	{\bibinfo {title} {{Nonlinear current-voltage behavior and giant positive
				magnetoresistance in nonmagnetic Au/Yttria-stabilized zirconia/Si
				heterostructures}},}\ }\href@noop {} {\bibfield  {journal} {\bibinfo
		{journal} {{Appl. Phys. Lett.}}\ }\textbf {\bibinfo {volume} {{95}}},\
	\bibinfo {pages} {{242113}} (\bibinfo {year} {{2009}})}\BibitemShut {NoStop}%
\bibitem [{\citenamefont {Nagahama}\ \emph {et~al.}(2007)\citenamefont
	{Nagahama}, \citenamefont {Santos},\ and\ \citenamefont {Moodera}}]{Nag07}%
\BibitemOpen
\bibfield  {author} {\bibinfo {author} {\bibfnamefont {T.}~\bibnamefont
		{Nagahama}}, \bibinfo {author} {\bibfnamefont {T.S.}\ \bibnamefont {Santos}},
	\ and\ \bibinfo {author} {\bibfnamefont {J.S.}\ \bibnamefont {Moodera}},\
}\bibfield  {title} {\enquote {\bibinfo {title} {{Enhanced magnetotransport
			at high bias in quasimagnetic tunnel junctions with EuS spin-filter
			barriers}},}\ }\href@noop {} {\bibfield  {journal} {\bibinfo  {journal}
	{Phys. Rev. Lett.}\ }\textbf {\bibinfo {volume} {99}},\ \bibinfo {pages}
{016602} (\bibinfo {year} {2007})}\BibitemShut {NoStop}%
\bibitem [{\citenamefont {Kim}\ \emph {et~al.}({2005})\citenamefont {Kim},
	\citenamefont {Javey}, \citenamefont {Tu}, \citenamefont {Cao}, \citenamefont
	{Wang},\ and\ \citenamefont {Dai}}]{Kim05}%
\BibitemOpen
\bibfield  {author} {\bibinfo {author} {\bibfnamefont {W.}~\bibnamefont
		{Kim}}, \bibinfo {author} {\bibfnamefont {A.}~\bibnamefont {Javey}}, \bibinfo
	{author} {\bibfnamefont {R.}~\bibnamefont {Tu}}, \bibinfo {author}
	{\bibfnamefont {J.}~\bibnamefont {Cao}}, \bibinfo {author} {\bibfnamefont
		{Q.}~\bibnamefont {Wang}}, \ and\ \bibinfo {author} {\bibfnamefont {H.J.}\
		\bibnamefont {Dai}},\ }\bibfield  {title} {\enquote {\bibinfo {title}
		{{Electrical contacts to carbon nanotubes down to 1 nm in diameter}},}\
}\href@noop {} {\bibfield  {journal} {\bibinfo  {journal} {{Appl. Phys.
			Lett.}}\ }\textbf {\bibinfo {volume} {{87}}},\ \bibinfo {pages} {173101}
(\bibinfo {year} {{2005}})}\BibitemShut {NoStop}%
\bibitem [{\citenamefont {Bockrath}\ \emph {et~al.}(1997)\citenamefont
	{Bockrath}, \citenamefont {Cobden}, \citenamefont {McEuen}, \citenamefont
	{Chopra}, \citenamefont {Zettl}, \citenamefont {Thess},\ and\ \citenamefont
	{Smalley}}]{Bockrath97}%
\BibitemOpen
\bibfield  {author} {\bibinfo {author} {\bibfnamefont {M.}~\bibnamefont
		{Bockrath}}, \bibinfo {author} {\bibfnamefont {D.H.}\ \bibnamefont {Cobden}},
	\bibinfo {author} {\bibfnamefont {P.L.}\ \bibnamefont {McEuen}}, \bibinfo
	{author} {\bibfnamefont {N.G.}\ \bibnamefont {Chopra}}, \bibinfo {author}
	{\bibfnamefont {A.}~\bibnamefont {Zettl}}, \bibinfo {author} {\bibfnamefont
		{A.}~\bibnamefont {Thess}}, \ and\ \bibinfo {author} {\bibfnamefont {R.E.}\
		\bibnamefont {Smalley}},\ }\bibfield  {title} {\enquote {\bibinfo {title}
		{Single-electron transport in ropes of carbon nanotubes},}\ }\href@noop {}
{\bibfield  {journal} {\bibinfo  {journal} {Science}\ }\textbf {\bibinfo
		{volume} {275}},\ \bibinfo {pages} {1922--1925} (\bibinfo {year}
	{1997})}\BibitemShut {NoStop}%
\bibitem [{\citenamefont {Nemec}\ \emph {et~al.}(2006)\citenamefont {Nemec},
	\citenamefont {Tom\'anek},\ and\ \citenamefont {Cuniberti}}]{Nemec06}%
\BibitemOpen
\bibfield  {author} {\bibinfo {author} {\bibfnamefont {N.}~\bibnamefont
		{Nemec}}, \bibinfo {author} {\bibfnamefont {D.}~\bibnamefont {Tom\'anek}}, \
	and\ \bibinfo {author} {\bibfnamefont {G.}~\bibnamefont {Cuniberti}},\
}\bibfield  {title} {\enquote {\bibinfo {title} {Contact dependence of
		carrier injection in carbon nanotubes: An ab initio study},}\ }\href@noop {}
{\bibfield  {journal} {\bibinfo  {journal} {Phys. Rev. Lett.}\ }\textbf
	{\bibinfo {volume} {96}},\ \bibinfo {pages} {076802} (\bibinfo {year}
	{2006})}\BibitemShut {NoStop}%
\bibitem [{\citenamefont {Shan}\ and\ \citenamefont {Cho}(2004)}]{Shan04}%
\BibitemOpen
\bibfield  {author} {\bibinfo {author} {\bibfnamefont {B.}~\bibnamefont
		{Shan}}\ and\ \bibinfo {author} {\bibfnamefont {K.}~\bibnamefont {Cho}},\
}\bibfield  {title} {\enquote {\bibinfo {title} {Ab initio study of
		\protect{Schottky} barriers at metal-nanotube contacts},}\ }\href@noop {}
{\bibfield  {journal} {\bibinfo  {journal} {Phys. Rev. B}\ }\textbf {\bibinfo
		{volume} {70}},\ \bibinfo {pages} {233405} (\bibinfo {year}
	{2004})}\BibitemShut {NoStop}%
\bibitem [{\citenamefont {Slonczewski}(1989)}]{Slonczewski89}%
\BibitemOpen
\bibfield  {author} {\bibinfo {author} {\bibfnamefont {J.C.}\ \bibnamefont
		{Slonczewski}},\ }\bibfield  {title} {\enquote {\bibinfo {title} {Conductance
			and exchange coupling of two ferromagnets separated by a tunneling
			barrier},}\ }\href@noop {} {\bibfield  {journal} {\bibinfo  {journal} {Phys.
			Rev. B}\ }\textbf {\bibinfo {volume} {39}},\ \bibinfo {pages} {6995--7002}
	(\bibinfo {year} {1989})}\BibitemShut {NoStop}%
\bibitem [{\citenamefont {Kane}\ and\ \citenamefont
	{Mele}(1997)}]{Kane_Phys.Rev.Lett.78/1997}%
\BibitemOpen
\bibfield  {author} {\bibinfo {author} {\bibfnamefont {C.L.}\ \bibnamefont
		{Kane}}\ and\ \bibinfo {author} {\bibfnamefont {E.J.}\ \bibnamefont {Mele}},\
}\bibfield  {title} {\enquote {\bibinfo {title} {Size, shape, and low energy
		electronic structure of carbon nanotubes},}\ }\href@noop {} {\bibfield
{journal} {\bibinfo  {journal} {Phys. Rev. Lett.}\ }\textbf {\bibinfo
	{volume} {78}},\ \bibinfo {pages} {1932} (\bibinfo {year}
{1997})}\BibitemShut {NoStop}%
\bibitem [{\citenamefont {Egger}\ and\ \citenamefont
	{Gogolin}(1998)}]{Egger_Eur.Phys.J.B3/1998}%
\BibitemOpen
\bibfield  {author} {\bibinfo {author} {\bibfnamefont {R.}~\bibnamefont
		{Egger}}\ and\ \bibinfo {author} {\bibfnamefont {A.O.}\ \bibnamefont
		{Gogolin}},\ }\bibfield  {title} {\enquote {\bibinfo {title} {Correlated
			transport and non-fermi-liquid behavior in single-wall carbon nanotubes},}\
}\href@noop {} {\bibfield  {journal} {\bibinfo  {journal} {Eur. Phys. J. B}\
}\textbf {\bibinfo {volume} {3}},\ \bibinfo {pages} {281--300} (\bibinfo
{year} {1998})}\BibitemShut {NoStop}%
\bibitem [{\citenamefont {Cottet}\ \emph
	{et~al.}(2006{\natexlab{a}})\citenamefont {Cottet}, \citenamefont {Kontos},
	\citenamefont {Sahoo}, \citenamefont {Man}, \citenamefont {Choi},
	\citenamefont {Belzig}, \citenamefont {Bruder}, \citenamefont {Morpurgo},\
	and\ \citenamefont {Sch{\"o}nenberger}}]{Cottet06}%
\BibitemOpen
\bibfield  {author} {\bibinfo {author} {\bibfnamefont {A.}~\bibnamefont
		{Cottet}}, \bibinfo {author} {\bibfnamefont {T.}~\bibnamefont {Kontos}},
	\bibinfo {author} {\bibfnamefont {S.}~\bibnamefont {Sahoo}}, \bibinfo
	{author} {\bibfnamefont {H.T.}\ \bibnamefont {Man}}, \bibinfo {author}
	{\bibfnamefont {M.-S.}\ \bibnamefont {Choi}}, \bibinfo {author}
	{\bibfnamefont {W.}~\bibnamefont {Belzig}}, \bibinfo {author} {\bibfnamefont
		{C.}~\bibnamefont {Bruder}}, \bibinfo {author} {\bibfnamefont {A.F.}\
		\bibnamefont {Morpurgo}}, \ and\ \bibinfo {author} {\bibfnamefont
		{C.}~\bibnamefont {Sch{\"o}nenberger}},\ }\bibfield  {title} {\enquote
	{\bibinfo {title} {Nanospintronics with carbon nanotubes},}\ }\href@noop {}
{\bibfield  {journal} {\bibinfo  {journal} {Sem. Sci. Tech.}\ }\textbf
	{\bibinfo {volume} {21}},\ \bibinfo {pages} {S78} (\bibinfo {year}
	{2006}{\natexlab{a}})}\BibitemShut {NoStop}%
\bibitem [{\citenamefont {Liang}\ \emph {et~al.}(2001)\citenamefont {Liang},
	\citenamefont {Bockrath}, \citenamefont {Bozovic}, \citenamefont {Hafner},
	\citenamefont {Tinkham},\ and\ \citenamefont {Park}}]{Liang_Nature411/2001}%
\BibitemOpen
\bibfield  {author} {\bibinfo {author} {\bibfnamefont {W.}~\bibnamefont
		{Liang}}, \bibinfo {author} {\bibfnamefont {M.}~\bibnamefont {Bockrath}},
	\bibinfo {author} {\bibfnamefont {D.}~\bibnamefont {Bozovic}}, \bibinfo
	{author} {\bibfnamefont {J.H.}\ \bibnamefont {Hafner}}, \bibinfo {author}
	{\bibfnamefont {M.}~\bibnamefont {Tinkham}}, \ and\ \bibinfo {author}
	{\bibfnamefont {H.}~\bibnamefont {Park}},\ }\bibfield  {title} {\enquote
	{\bibinfo {title} {{Fabry-P\'erot interference in a nanotube electron
				waveguide}},}\ }\href@noop {} {\bibfield  {journal} {\bibinfo  {journal}
		{Nature}\ }\textbf {\bibinfo {volume} {411}},\ \bibinfo {pages} {665--669}
	(\bibinfo {year} {2001})}\BibitemShut {NoStop}%
\bibitem [{\citenamefont {Man}\ and\ \citenamefont {Morpurgo}(2005)}]{Man05}%
\BibitemOpen
\bibfield  {author} {\bibinfo {author} {\bibfnamefont {H.T.}\ \bibnamefont
		{Man}}\ and\ \bibinfo {author} {\bibfnamefont {A.F.}\ \bibnamefont
		{Morpurgo}},\ }\bibfield  {title} {\enquote {\bibinfo {title}
		{Sample-specific and ensemble-averaged magnetoconductance of individual
			single-wall carbon nanotubes},}\ }\href@noop {} {\bibfield  {journal}
	{\bibinfo  {journal} {Phys. Rev. Lett.}\ }\textbf {\bibinfo {volume} {95}},\
	\bibinfo {pages} {026801} (\bibinfo {year} {2005})}\BibitemShut {NoStop}%
\bibitem [{\citenamefont {Cottet}\ \emph
	{et~al.}(2006{\natexlab{b}})\citenamefont {Cottet}, \citenamefont {Kontos},
	\citenamefont {Belzig}, \citenamefont {Sch\"{o}nenberger},\ and\
	\citenamefont {Bruder}}]{Cottet_Europhys.Lett.74/2006}%
\BibitemOpen
\bibfield  {author} {\bibinfo {author} {\bibfnamefont {A.}~\bibnamefont
		{Cottet}}, \bibinfo {author} {\bibfnamefont {T.}~\bibnamefont {Kontos}},
	\bibinfo {author} {\bibfnamefont {W.}~\bibnamefont {Belzig}}, \bibinfo
	{author} {\bibfnamefont {C.}~\bibnamefont {Sch\"{o}nenberger}}, \ and\
	\bibinfo {author} {\bibfnamefont {C.}~\bibnamefont {Bruder}},\ }\bibfield
{title} {\enquote {\bibinfo {title} {Controlling spin in an electronic
			interferometer with spin-active interfaces},}\ }\href@noop {} {\bibfield
	{journal} {\bibinfo  {journal} {Europhys. Lett.}\ }\textbf {\bibinfo {volume}
		{74}},\ \bibinfo {pages} {320--326} (\bibinfo {year}
	{2006}{\natexlab{b}})}\BibitemShut {NoStop}%
\bibitem [{\citenamefont {Blonder}\ \emph {et~al.}(1982)\citenamefont
	{Blonder}, \citenamefont {Tinkham},\ and\ \citenamefont
	{Klapwijk}}]{Blonder_Phys.Rev.B25/1982}%
\BibitemOpen
\bibfield  {author} {\bibinfo {author} {\bibfnamefont {G.E.}\ \bibnamefont
		{Blonder}}, \bibinfo {author} {\bibfnamefont {M.}~\bibnamefont {Tinkham}}, \
	and\ \bibinfo {author} {\bibfnamefont {T.M.}\ \bibnamefont {Klapwijk}},\
}\bibfield  {title} {\enquote {\bibinfo {title} {Transition from metallic to
		tunneling regimes in superconducting microconstrictions: Excess current,
		charge imbalance, and supercurrent conversion},}\ }\href@noop {} {\bibfield
{journal} {\bibinfo  {journal} {Phys. Rev. B}\ }\textbf {\bibinfo {volume}
	{25}},\ \bibinfo {pages} {4515--4532} (\bibinfo {year} {1982})}\BibitemShut
{NoStop}%
\bibitem [{\citenamefont {Qi}\ \emph {et~al.}(1998)\citenamefont {Qi},
	\citenamefont {Xing},\ and\ \citenamefont {Dong}}]{Qi_Phys.Rev.B58/1998}%
\BibitemOpen
\bibfield  {author} {\bibinfo {author} {\bibfnamefont {Y.}~\bibnamefont
		{Qi}}, \bibinfo {author} {\bibfnamefont {D.Y.}\ \bibnamefont {Xing}}, \ and\
	\bibinfo {author} {\bibfnamefont {J.}~\bibnamefont {Dong}},\ }\bibfield
{title} {\enquote {\bibinfo {title} {{Relation between Julliere and
				Slonczewski models of tunneling magnetoresistance}},}\ }\href@noop {}
{\bibfield  {journal} {\bibinfo  {journal} {Phys. Rev. B}\ }\textbf {\bibinfo
		{volume} {58}},\ \bibinfo {pages} {2783--2787} (\bibinfo {year}
	{1998})}\BibitemShut {NoStop}%
\bibitem [{\citenamefont {Grundler}(2001)}]{Grundler_Phys.Rev.Lett.86/2001}%
\BibitemOpen
\bibfield  {author} {\bibinfo {author} {\bibfnamefont {D.}~\bibnamefont
		{Grundler}},\ }\bibfield  {title} {\enquote {\bibinfo {title} {Oscillatory
			spin-filtering due to gate control of spin-dependent interface
			conductance},}\ }\href@noop {} {\bibfield  {journal} {\bibinfo  {journal}
		{Phys. Rev. Lett.}\ }\textbf {\bibinfo {volume} {86}},\ \bibinfo {pages}
	{1058} (\bibinfo {year} {2001})}\BibitemShut {NoStop}%
\bibitem [{\citenamefont {Hu}\ and\ \citenamefont
	{Matsuyama}(2001)}]{Hu_Phys.Rev.Lett.87/2001}%
\BibitemOpen
\bibfield  {author} {\bibinfo {author} {\bibfnamefont {C.-M.}\ \bibnamefont
		{Hu}}\ and\ \bibinfo {author} {\bibfnamefont {T.}~\bibnamefont {Matsuyama}},\
}\bibfield  {title} {\enquote {\bibinfo {title} {Spin injection across a
		heterojunction: A ballistic picture},}\ }\href@noop {} {\bibfield  {journal}
{\bibinfo  {journal} {Phys. Rev. Lett.}\ }\textbf {\bibinfo {volume} {87}},\
\bibinfo {pages} {066803} (\bibinfo {year} {2001})}\BibitemShut {NoStop}%
\bibitem [{\citenamefont {Datta}(1997)}]{Datta_book}%
\BibitemOpen
\bibfield  {author} {\bibinfo {author} {\bibfnamefont {S.}~\bibnamefont
		{Datta}},\ }\href@noop {} {\emph {\bibinfo {title} {Electronic transport in
			mesoscopic systems}}}\ (\bibinfo  {publisher} {Cambridge University Press},\
\bibinfo {address} {Cambridge},\ \bibinfo {year} {1997})\BibitemShut
{NoStop}%
\bibitem [{\citenamefont {Blanter}\ and\ \citenamefont
	{B{\"u}ttiker}(2000)}]{Blanter_Phys.Rep.336/2000}%
\BibitemOpen
\bibfield  {author} {\bibinfo {author} {\bibfnamefont {Ya.M.}\ \bibnamefont
		{Blanter}}\ and\ \bibinfo {author} {\bibfnamefont {M.}~\bibnamefont
		{B{\"u}ttiker}},\ }\bibfield  {title} {\enquote {\bibinfo {title} {Shot noise
			in mesoscopic conductors},}\ }\href@noop {} {\bibfield  {journal} {\bibinfo
		{journal} {Phys. Rep.}\ }\textbf {\bibinfo {volume} {336}},\ \bibinfo {pages}
	{1--166} (\bibinfo {year} {2000})}\BibitemShut {NoStop}%
\bibitem [{\citenamefont {McEuen}\ \emph {et~al.}(1999)\citenamefont {McEuen},
	\citenamefont {Bockrath}, \citenamefont {Cobden}, \citenamefont {Yoon},\ and\
	\citenamefont {Louie}}]{McEuen_Phys.Rev.Lett.83/1999}%
\BibitemOpen
\bibfield  {author} {\bibinfo {author} {\bibfnamefont {P.L.}\ \bibnamefont
		{McEuen}}, \bibinfo {author} {\bibfnamefont {M.}~\bibnamefont {Bockrath}},
	\bibinfo {author} {\bibfnamefont {D.H.}\ \bibnamefont {Cobden}}, \bibinfo
	{author} {\bibfnamefont {Y.-G.}\ \bibnamefont {Yoon}}, \ and\ \bibinfo
	{author} {\bibfnamefont {S.G.}\ \bibnamefont {Louie}},\ }\bibfield  {title}
{\enquote {\bibinfo {title} {Disorder, pseudospins, and backscattering in
			carbon nanotubes},}\ }\href@noop {} {\bibfield  {journal} {\bibinfo
		{journal} {Phys. Rev. Lett.}\ }\textbf {\bibinfo {volume} {83}},\ \bibinfo
	{pages} {5098} (\bibinfo {year} {1999})}\BibitemShut {NoStop}%
\bibitem [{\citenamefont {Veillette}\ \emph {et~al.}(2004)\citenamefont
	{Veillette}, \citenamefont {Bena},\ and\ \citenamefont
	{Balents}}]{Veillette_Phys.Rev.B69/2004}%
\BibitemOpen
\bibfield  {author} {\bibinfo {author} {\bibfnamefont {M.Y.}\ \bibnamefont
		{Veillette}}, \bibinfo {author} {\bibfnamefont {C.}~\bibnamefont {Bena}}, \
	and\ \bibinfo {author} {\bibfnamefont {L.}~\bibnamefont {Balents}},\
}\bibfield  {title} {\enquote {\bibinfo {title} {Spin precession and
		oscillations in mesoscopic systems},}\ }\href@noop {} {\bibfield  {journal}
{\bibinfo  {journal} {Phys. Rev. B}\ }\textbf {\bibinfo {volume} {69}},\
\bibinfo {pages} {075319} (\bibinfo {year} {2004})}\BibitemShut {NoStop}%
\bibitem [{\citenamefont {Misiorny}\ and\ \citenamefont
	{Meyer}(2016)}]{TheoryPaper}%
\BibitemOpen
\bibfield  {author} {\bibinfo {author} {\bibfnamefont {M.}~\bibnamefont
		{Misiorny}}\ and\ \bibinfo {author} {\bibfnamefont {C}~\bibnamefont
		{Meyer}},\ }\href@noop {} {\bibfield  {journal} {\bibinfo  {journal} {in
			preparation}\ } (\bibinfo {year} {2016})}\BibitemShut {NoStop}%
\bibitem [{\citenamefont {Tedrow}\ and\ \citenamefont
	{Meservey}(1973)}]{Tedrow73}%
\BibitemOpen
\bibfield  {author} {\bibinfo {author} {\bibfnamefont {P.M.}\ \bibnamefont
		{Tedrow}}\ and\ \bibinfo {author} {\bibfnamefont {R.}~\bibnamefont
		{Meservey}},\ }\bibfield  {title} {\enquote {\bibinfo {title} {{Spin
				polarization of electrons tunneling from films of Fe, Co, Ni, and Gd}},}\
}\href@noop {} {\bibfield  {journal} {\bibinfo  {journal} {Phys. Rev. B}\
}\textbf {\bibinfo {volume} {7}},\ \bibinfo {pages} {318--326} (\bibinfo
{year} {1973})}\BibitemShut {NoStop}%
\bibitem [{\citenamefont {Shan}\ \emph {et~al.}(1993)\citenamefont {Shan},
	\citenamefont {He}, \citenamefont {Moore}, \citenamefont {Woollam},\ and\
	\citenamefont {Sellmyer}}]{Shan93}%
\BibitemOpen
\bibfield  {author} {\bibinfo {author} {\bibfnamefont {Z.-S.}\ \bibnamefont
		{Shan}}, \bibinfo {author} {\bibfnamefont {Ping}\ \bibnamefont {He}},
	\bibinfo {author} {\bibfnamefont {C.}~\bibnamefont {Moore}}, \bibinfo
	{author} {\bibfnamefont {John}\ \bibnamefont {Woollam}}, \ and\ \bibinfo
	{author} {\bibfnamefont {D.~J.}\ \bibnamefont {Sellmyer}},\ }\bibfield
{title} {\enquote {\bibinfo {title} {{Behavior of disordered Co-Pd, Co-Ag,
				and Co-Mo alloys in multilayer interfaces}},}\ }\href@noop {} {\bibfield
	{journal} {\bibinfo  {journal} {J. Appl. Phys.}\ }\textbf {\bibinfo {volume}
		{73}},\ \bibinfo {pages} {6057--6059} (\bibinfo {year} {1993})}\BibitemShut
{NoStop}%
\bibitem [{\citenamefont {Ranjan}\ \emph {et~al.}(2004)\citenamefont {Ranjan},
	\citenamefont {Guti\'errez}, \citenamefont {Krompiewski},\ and\ \citenamefont
	{Cuniberti}}]{Ranjan04}%
\BibitemOpen
\bibfield  {author} {\bibinfo {author} {\bibfnamefont {N.}~\bibnamefont
		{Ranjan}}, \bibinfo {author} {\bibfnamefont {R.}~\bibnamefont {Guti\'errez}},
	\bibinfo {author} {\bibfnamefont {S.}~\bibnamefont {Krompiewski}}, \ and\
	\bibinfo {author} {\bibfnamefont {G.}~\bibnamefont {Cuniberti}},\ }\bibfield
{title} {\enquote {\bibinfo {title} {Electron transport in carbon
			nanotube-metal system: contact effects},}\ }\href@noop {} {\bibfield
	{journal} {\bibinfo  {journal} {Mol. Phys. Rep.}\ }\textbf {\bibinfo {volume}
		{40}},\ \bibinfo {pages} {125--130} (\bibinfo {year} {2004})}\BibitemShut
{NoStop}%
\bibitem [{\citenamefont {Yuasa}\ \emph {et~al.}(2004)\citenamefont {Yuasa},
	\citenamefont {Nagahama}, \citenamefont {Fukushima}, \citenamefont {Suzuki},\
	and\ \citenamefont {Ando}}]{Yuasa04}%
\BibitemOpen
\bibfield  {author} {\bibinfo {author} {\bibfnamefont {S.}~\bibnamefont
		{Yuasa}}, \bibinfo {author} {\bibfnamefont {T.}~\bibnamefont {Nagahama}},
	\bibinfo {author} {\bibfnamefont {A.}~\bibnamefont {Fukushima}}, \bibinfo
	{author} {\bibfnamefont {Y.}~\bibnamefont {Suzuki}}, \ and\ \bibinfo {author}
	{\bibfnamefont {K.}~\bibnamefont {Ando}},\ }\bibfield  {title} {\enquote
	{\bibinfo {title} {{Giant room-temperature magnetoresistance in
				single-crystal Fe/MgO/Fe magnetic tunnel junctions}},}\ }\href@noop {}
{\bibfield  {journal} {\bibinfo  {journal} {Nat. Mater.}\ }\textbf {\bibinfo
		{volume} {{3}}},\ \bibinfo {pages} {{868--871}} (\bibinfo {year}
	{2004})}\BibitemShut {NoStop}%
\bibitem [{\citenamefont {Marukame}\ \emph {et~al.}(2010)\citenamefont
	{Marukame}, \citenamefont {Ishikawa}, \citenamefont {Taira}, \citenamefont
	{Matsuda}, \citenamefont {Uemura},\ and\ \citenamefont
	{Yamamoto}}]{Marukame10}%
\BibitemOpen
\bibfield  {author} {\bibinfo {author} {\bibfnamefont {T.}~\bibnamefont
		{Marukame}}, \bibinfo {author} {\bibfnamefont {T.}~\bibnamefont {Ishikawa}},
	\bibinfo {author} {\bibfnamefont {T.}~\bibnamefont {Taira}}, \bibinfo
	{author} {\bibfnamefont {K.}~\bibnamefont {Matsuda}}, \bibinfo {author}
	{\bibfnamefont {T.}~\bibnamefont {Uemura}}, \ and\ \bibinfo {author}
	{\bibfnamefont {M.}~\bibnamefont {Yamamoto}},\ }\bibfield  {title} {\enquote
	{\bibinfo {title} {{Giant oscillations in spin-dependent tunneling
				resistances as a function of barrier thickness in fully epitaxial magnetic
				tunnel junctions with a MgO barrier}},}\ }\href@noop {} {\bibfield  {journal}
	{\bibinfo  {journal} {Phys. Rev. B}\ }\textbf {\bibinfo {volume} {81}},\
	\bibinfo {pages} {134432} (\bibinfo {year} {2010})}\BibitemShut {NoStop}%
\bibitem [{\citenamefont {Wang}\ \emph {et~al.}(2010)\citenamefont {Wang},
	\citenamefont {Liu}, \citenamefont {Kodzuka}, \citenamefont {Sukegawa},
	\citenamefont {Wojcik}, \citenamefont {Jedryka}, \citenamefont {Wu},
	\citenamefont {Inomata}, \citenamefont {Mitani},\ and\ \citenamefont
	{Hono}}]{Wang10}%
\BibitemOpen
\bibfield  {author} {\bibinfo {author} {\bibfnamefont {W.}~\bibnamefont
		{Wang}}, \bibinfo {author} {\bibfnamefont {E.}~\bibnamefont {Liu}}, \bibinfo
	{author} {\bibfnamefont {M.}~\bibnamefont {Kodzuka}}, \bibinfo {author}
	{\bibfnamefont {H.}~\bibnamefont {Sukegawa}}, \bibinfo {author}
	{\bibfnamefont {M.}~\bibnamefont {Wojcik}}, \bibinfo {author} {\bibfnamefont
		{E.}~\bibnamefont {Jedryka}}, \bibinfo {author} {\bibfnamefont {G.~H.}\
		\bibnamefont {Wu}}, \bibinfo {author} {\bibfnamefont {K.}~\bibnamefont
		{Inomata}}, \bibinfo {author} {\bibfnamefont {S.}~\bibnamefont {Mitani}}, \
	and\ \bibinfo {author} {\bibfnamefont {K.}~\bibnamefont {Hono}},\ }\bibfield
{title} {\enquote {\bibinfo {title} {{Coherent tunneling and giant tunneling
				magnetoresistance in Co$_2$FeAl/MgO/CoFe magnetic tunneling junctions}},}\
}\href@noop {} {\bibfield  {journal} {\bibinfo  {journal} {Phys. Rev. B}\
}\textbf {\bibinfo {volume} {81}},\ \bibinfo {pages} {140402} (\bibinfo
{year} {2010})}\BibitemShut {NoStop}%
\bibitem [{\citenamefont {Butler}\ \emph {et~al.}(2001)\citenamefont {Butler},
	\citenamefont {Zhang}, \citenamefont {Schulthess},\ and\ \citenamefont
	{MacLaren}}]{Butler01}%
\BibitemOpen
\bibfield  {author} {\bibinfo {author} {\bibfnamefont {W.~H.}\ \bibnamefont
		{Butler}}, \bibinfo {author} {\bibfnamefont {X.-G.}\ \bibnamefont {Zhang}},
	\bibinfo {author} {\bibfnamefont {T.~C.}\ \bibnamefont {Schulthess}}, \ and\
	\bibinfo {author} {\bibfnamefont {J.~M.}\ \bibnamefont {MacLaren}},\
}\bibfield  {title} {\enquote {\bibinfo {title} {{Spin-dependent tunneling
			conductance of $\mathrm{Fe}|\mathrm{MgO}|\mathrm{Fe}$ sandwiches}},}\
}\href@noop {} {\bibfield  {journal} {\bibinfo  {journal} {Phys. Rev. B}\
}\textbf {\bibinfo {volume} {63}},\ \bibinfo {pages} {054416} (\bibinfo
{year} {2001})}\BibitemShut {NoStop}%
\bibitem [{\citenamefont {Aurich}\ \emph {et~al.}(2010)\citenamefont {Aurich},
	\citenamefont {Baumgartner}, \citenamefont {Freitag}, \citenamefont
	{Eichler}, \citenamefont {Trbovic},\ and\ \citenamefont
	{Sch\"onenberger}}]{Aurich10}%
\BibitemOpen
\bibfield  {author} {\bibinfo {author} {\bibfnamefont {H.}~\bibnamefont
		{Aurich}}, \bibinfo {author} {\bibfnamefont {A.}~\bibnamefont {Baumgartner}},
	\bibinfo {author} {\bibfnamefont {F.}~\bibnamefont {Freitag}}, \bibinfo
	{author} {\bibfnamefont {A.}~\bibnamefont {Eichler}}, \bibinfo {author}
	{\bibfnamefont {J.}~\bibnamefont {Trbovic}}, \ and\ \bibinfo {author}
	{\bibfnamefont {C.}~\bibnamefont {Sch\"onenberger}},\ }\bibfield  {title}
{\enquote {\bibinfo {title} {Permalloy-based carbon nanotube spin-valve},}\
}\href@noop {} {\bibfield  {journal} {\bibinfo  {journal} {App. Phys. Lett.}\
}\textbf {\bibinfo {volume} {97}},\ \bibinfo {pages} {153116} (\bibinfo
{year} {2010})}\BibitemShut {NoStop}%
\bibitem [{\citenamefont {Zhang}\ and\ \citenamefont
	{Levy}(1999)}]{Zhang_Eur.Phys.J.B10/1999}%
\BibitemOpen
\bibfield  {author} {\bibinfo {author} {\bibfnamefont {S.}~\bibnamefont
		{Zhang}}\ and\ \bibinfo {author} {\bibfnamefont {P.M.}\ \bibnamefont
		{Levy}},\ }\bibfield  {title} {\enquote {\bibinfo {title} {Models for
			magnetoresistance in tunnel junctions},}\ }\href@noop {} {\bibfield
	{journal} {\bibinfo  {journal} {Eur. Phys. J. B}\ }\textbf {\bibinfo {volume}
		{10}},\ \bibinfo {pages} {599--606} (\bibinfo {year} {1999})}\BibitemShut
{NoStop}%
\bibitem [{\citenamefont {Mavropoulos}\ \emph {et~al.}(2004)\citenamefont
	{Mavropoulos}, \citenamefont {Papanikolaou},\ and\ \citenamefont
	{Dederichs}}]{Mavropoulos_Phys.Rev.B69/2004}%
\BibitemOpen
\bibfield  {author} {\bibinfo {author} {\bibfnamefont {P.}~\bibnamefont
		{Mavropoulos}}, \bibinfo {author} {\bibfnamefont {N.}~\bibnamefont
		{Papanikolaou}}, \ and\ \bibinfo {author} {\bibfnamefont {P.H.}\ \bibnamefont
		{Dederichs}},\ }\bibfield  {title} {\enquote {\bibinfo {title}
		{{Korringa-Kohn-Rostoker Green-function formalism for ballistic
				transport}},}\ }\href@noop {} {\bibfield  {journal} {\bibinfo  {journal}
		{Phys. Rev. B}\ }\textbf {\bibinfo {volume} {69}},\ \bibinfo {pages} {125104}
	(\bibinfo {year} {2004})}\BibitemShut {NoStop}%
\bibitem [{\citenamefont {Tombros}\ \emph {et~al.}(2006)\citenamefont
	{Tombros}, \citenamefont {van~der Molen},\ and\ \citenamefont {van
		Wees}}]{Tombros06}%
\BibitemOpen
\bibfield  {author} {\bibinfo {author} {\bibfnamefont {N.}~\bibnamefont
		{Tombros}}, \bibinfo {author} {\bibfnamefont {S.J.}\ \bibnamefont {van~der
			Molen}}, \ and\ \bibinfo {author} {\bibfnamefont {B.J.}\ \bibnamefont {van
			Wees}},\ }\bibfield  {title} {\enquote {\bibinfo {title} {Separating spin and
			charge transport in single-wall carbon nanotubes},}\ }\href@noop {}
{\bibfield  {journal} {\bibinfo  {journal} {Phys. Rev. B}\ }\textbf {\bibinfo
		{volume} {73}},\ \bibinfo {pages} {233403} (\bibinfo {year}
	{2006})}\BibitemShut {NoStop}%
\bibitem [{\citenamefont {Ono}\ \emph {et~al.}(1998)\citenamefont {Ono},
	\citenamefont {Shimada},\ and\ \citenamefont {Ootuka}}]{Ono98}%
\BibitemOpen
\bibfield  {author} {\bibinfo {author} {\bibfnamefont {K.}~\bibnamefont
		{Ono}}, \bibinfo {author} {\bibfnamefont {H.}~\bibnamefont {Shimada}}, \ and\
	\bibinfo {author} {\bibfnamefont {Y.}~\bibnamefont {Ootuka}},\ }\bibfield
{title} {\enquote {\bibinfo {title} {Spin polarization and
			\protect{Magneto-Coulomb} oscillations in ferromagnetic single electron
			devices},}\ }\href@noop {} {\bibfield  {journal} {\bibinfo  {journal} {J.
			Phys. Soc. Jpn.}\ }\textbf {\bibinfo {volume} {67}},\ \bibinfo {pages}
	{2852--2856} (\bibinfo {year} {1998})}\BibitemShut {NoStop}%
\bibitem [{\citenamefont {Shimada}\ \emph {et~al.}(2003)\citenamefont
	{Shimada}, \citenamefont {Ono},\ and\ \citenamefont {Ootuka}}]{Shi03}%
\BibitemOpen
\bibfield  {author} {\bibinfo {author} {\bibfnamefont {H.}~\bibnamefont
		{Shimada}}, \bibinfo {author} {\bibfnamefont {K.}~\bibnamefont {Ono}}, \ and\
	\bibinfo {author} {\bibfnamefont {Y.}~\bibnamefont {Ootuka}},\ }\bibfield
{title} {\enquote {\bibinfo {title} {Driving the single-electron device with
			a magnetic field (invited)},}\ }\href@noop {} {\bibfield  {journal} {\bibinfo
		{journal} {J. Appl. Phys.}\ }\textbf {\bibinfo {volume} {93}},\ \bibinfo
	{pages} {8259--8264} (\bibinfo {year} {2003})}\BibitemShut {NoStop}%
\bibitem [{\citenamefont {van~der Molen}\ \emph {et~al.}(2006)\citenamefont
	{van~der Molen}, \citenamefont {Tombros},\ and\ \citenamefont {van
		Wees}}]{Sense06}%
\BibitemOpen
\bibfield  {author} {\bibinfo {author} {\bibfnamefont {S.J.}\ \bibnamefont
		{van~der Molen}}, \bibinfo {author} {\bibfnamefont {N.}~\bibnamefont
		{Tombros}}, \ and\ \bibinfo {author} {\bibfnamefont {B.J.}\ \bibnamefont {van
			Wees}},\ }\bibfield  {title} {\enquote {\bibinfo {title} {Magneto-coulomb
			effect in spin-valve devices},}\ }\href@noop {} {\bibfield  {journal}
	{\bibinfo  {journal} {Phys. Rev. B}\ }\textbf {\bibinfo {volume} {73}},\
	\bibinfo {pages} {220406} (\bibinfo {year} {2006})}\BibitemShut {NoStop}%
\bibitem [{\citenamefont {Gould}\ \emph {et~al.}(2004)\citenamefont {Gould},
	\citenamefont {R\"uster}, \citenamefont {Jungwirth}, \citenamefont {Girgis},
	\citenamefont {Schott}, \citenamefont {Giraud}, \citenamefont {Brunner},
	\citenamefont {Schmidt},\ and\ \citenamefont {Molenkamp}}]{Gould04}%
\BibitemOpen
\bibfield  {author} {\bibinfo {author} {\bibfnamefont {C.}~\bibnamefont
		{Gould}}, \bibinfo {author} {\bibfnamefont {C.}~\bibnamefont {R\"uster}},
	\bibinfo {author} {\bibfnamefont {T.}~\bibnamefont {Jungwirth}}, \bibinfo
	{author} {\bibfnamefont {E.}~\bibnamefont {Girgis}}, \bibinfo {author}
	{\bibfnamefont {G.M.}\ \bibnamefont {Schott}}, \bibinfo {author}
	{\bibfnamefont {R.}~\bibnamefont {Giraud}}, \bibinfo {author} {\bibfnamefont
		{K.}~\bibnamefont {Brunner}}, \bibinfo {author} {\bibfnamefont
		{G.}~\bibnamefont {Schmidt}}, \ and\ \bibinfo {author} {\bibfnamefont {L.W.}\
		\bibnamefont {Molenkamp}},\ }\bibfield  {title} {\enquote {\bibinfo {title}
		{Tunneling anisotropic magnetoresistance: A spin-valve-like tunnel
			magnetoresistance using a single magnetic layer},}\ }\href@noop {} {\bibfield
	{journal} {\bibinfo  {journal} {Phys. Rev. Lett.}\ }\textbf {\bibinfo
		{volume} {93}},\ \bibinfo {pages} {117203} (\bibinfo {year}
	{2004})}\BibitemShut {NoStop}%
\bibitem [{\citenamefont {Gr\"unewald}\ \emph {et~al.}(2011)\citenamefont
	{Gr\"unewald}, \citenamefont {Wahler}, \citenamefont {Schumann},
	\citenamefont {Michelfeit}, \citenamefont {Gould}, \citenamefont {Schmidt},
	\citenamefont {W\"urthner}, \citenamefont {Schmidt},\ and\ \citenamefont
	{Molenkamp}}]{Gru11}%
\BibitemOpen
\bibfield  {author} {\bibinfo {author} {\bibfnamefont {M.}~\bibnamefont
		{Gr\"unewald}}, \bibinfo {author} {\bibfnamefont {M.}~\bibnamefont {Wahler}},
	\bibinfo {author} {\bibfnamefont {F.}~\bibnamefont {Schumann}}, \bibinfo
	{author} {\bibfnamefont {M.}~\bibnamefont {Michelfeit}}, \bibinfo {author}
	{\bibfnamefont {C.}~\bibnamefont {Gould}}, \bibinfo {author} {\bibfnamefont
		{R.}~\bibnamefont {Schmidt}}, \bibinfo {author} {\bibfnamefont
		{F.}~\bibnamefont {W\"urthner}}, \bibinfo {author} {\bibfnamefont
		{G.}~\bibnamefont {Schmidt}}, \ and\ \bibinfo {author} {\bibfnamefont {L.W.}\
		\bibnamefont {Molenkamp}},\ }\bibfield  {title} {\enquote {\bibinfo {title}
		{Tunneling anisotropic magnetoresistance in organic spin valves},}\
}\href@noop {} {\bibfield  {journal} {\bibinfo  {journal} {Phys. Rev. B}\
}\textbf {\bibinfo {volume} {84}},\ \bibinfo {pages} {125208} (\bibinfo
{year} {2011})}\BibitemShut {NoStop}%
\bibitem [{\citenamefont {Wang}\ \emph {et~al.}(2013)\citenamefont {Wang},
	\citenamefont {Tran}, \citenamefont {Brinks}, \citenamefont {Sanderink},
	\citenamefont {Bolhuis}, \citenamefont {van~der Wiel},\ and\ \citenamefont
	{de~Jong}}]{Wang13}%
\BibitemOpen
\bibfield  {author} {\bibinfo {author} {\bibfnamefont {K.}~\bibnamefont
		{Wang}}, \bibinfo {author} {\bibfnamefont {T.L.A.}\ \bibnamefont {Tran}},
	\bibinfo {author} {\bibfnamefont {P.}~\bibnamefont {Brinks}}, \bibinfo
	{author} {\bibfnamefont {J.G.M.}\ \bibnamefont {Sanderink}}, \bibinfo
	{author} {\bibfnamefont {T.}~\bibnamefont {Bolhuis}}, \bibinfo {author}
	{\bibfnamefont {W.G.}\ \bibnamefont {van~der Wiel}}, \ and\ \bibinfo {author}
	{\bibfnamefont {M.P.}\ \bibnamefont {de~Jong}},\ }\bibfield  {title}
{\enquote {\bibinfo {title} {{Tunneling anisotropic magnetoresistance in
				Co/AlO${}_{x}$/Al tunnel junctions with fcc Co (111) electrodes}},}\
}\href@noop {} {\bibfield  {journal} {\bibinfo  {journal} {Phys. Rev. B}\
}\textbf {\bibinfo {volume} {88}},\ \bibinfo {pages} {054407} (\bibinfo
{year} {2013})}\BibitemShut {NoStop}%
\bibitem [{\citenamefont {Dash}\ \emph {et~al.}(2009)\citenamefont {Dash},
	\citenamefont {Sharma}, \citenamefont {Patel}, \citenamefont {de~Jong},\ and\
	\citenamefont {Jansen}}]{Dash09}%
\BibitemOpen
\bibfield  {author} {\bibinfo {author} {\bibfnamefont {S.P.}\ \bibnamefont
		{Dash}}, \bibinfo {author} {\bibfnamefont {S.}~\bibnamefont {Sharma}},
	\bibinfo {author} {\bibfnamefont {R.S.}\ \bibnamefont {Patel}}, \bibinfo
	{author} {\bibfnamefont {M.P.}\ \bibnamefont {de~Jong}}, \ and\ \bibinfo
	{author} {\bibfnamefont {R.}~\bibnamefont {Jansen}},\ }\bibfield  {title}
{\enquote {\bibinfo {title} {Electrical creation of spin polarization in
			silicon at room temperature},}\ }\href@noop {} {\bibfield  {journal}
	{\bibinfo  {journal} {Nature}\ }\textbf {\bibinfo {volume} {462}},\ \bibinfo
	{pages} {491--494} (\bibinfo {year} {2009})}\BibitemShut {NoStop}%
\bibitem [{\citenamefont {Cobden}\ \emph {et~al.}(1998)\citenamefont {Cobden},
	\citenamefont {Bockrath}, \citenamefont {McEuen}, \citenamefont {Rinzler},\
	and\ \citenamefont {Smalley}}]{Cobden98}%
\BibitemOpen
\bibfield  {author} {\bibinfo {author} {\bibfnamefont {D.H.}\ \bibnamefont
		{Cobden}}, \bibinfo {author} {\bibfnamefont {M.}~\bibnamefont {Bockrath}},
	\bibinfo {author} {\bibfnamefont {P.L.}\ \bibnamefont {McEuen}}, \bibinfo
	{author} {\bibfnamefont {A.G.}\ \bibnamefont {Rinzler}}, \ and\ \bibinfo
	{author} {\bibfnamefont {R.E.}\ \bibnamefont {Smalley}},\ }\bibfield  {title}
{\enquote {\bibinfo {title} {Spin splitting and even-odd effects in carbon
			nanotubes},}\ }\href@noop {} {\bibfield  {journal} {\bibinfo  {journal}
		{Phys. Rev. Lett.}\ }\textbf {\bibinfo {volume} {81}},\ \bibinfo {pages}
	{681--684} (\bibinfo {year} {1998})}\BibitemShut {NoStop}%
\bibitem [{\citenamefont {Han}\ and\ \citenamefont {Kawakami}(2011)}]{Han11}%
\BibitemOpen
\bibfield  {author} {\bibinfo {author} {\bibfnamefont {W.}~\bibnamefont
		{Han}}\ and\ \bibinfo {author} {\bibfnamefont {R.K.}\ \bibnamefont
		{Kawakami}},\ }\bibfield  {title} {\enquote {\bibinfo {title} {Spin
			relaxation in single-layer and bilayer graphene},}\ }\href@noop {} {\bibfield
	{journal} {\bibinfo  {journal} {Phys. Rev. Lett.}\ }\textbf {\bibinfo
		{volume} {107}},\ \bibinfo {pages} {047207} (\bibinfo {year}
	{2011})}\BibitemShut {NoStop}%
\bibitem [{\citenamefont {Avsar}\ \emph {et~al.}(2011)\citenamefont {Avsar},
	\citenamefont {Yang}, \citenamefont {Bae}, \citenamefont {Balakrishnan},
	\citenamefont {Volmer}, \citenamefont {Jaiswal}, \citenamefont {Yi},
	\citenamefont {Ali}, \citenamefont {G\"untherodt}, \citenamefont {Hong},
	\citenamefont {Beschoten},\ and\ \citenamefont {\"Ozyilmaz}}]{Avsar11}%
\BibitemOpen
\bibfield  {author} {\bibinfo {author} {\bibfnamefont {A.}~\bibnamefont
		{Avsar}}, \bibinfo {author} {\bibfnamefont {T.-Y.}\ \bibnamefont {Yang}},
	\bibinfo {author} {\bibfnamefont {S.}~\bibnamefont {Bae}}, \bibinfo {author}
	{\bibfnamefont {J.}~\bibnamefont {Balakrishnan}}, \bibinfo {author}
	{\bibfnamefont {F.}~\bibnamefont {Volmer}}, \bibinfo {author} {\bibfnamefont
		{M.}~\bibnamefont {Jaiswal}}, \bibinfo {author} {\bibfnamefont
		{Z.}~\bibnamefont {Yi}}, \bibinfo {author} {\bibfnamefont {S.R.}\
		\bibnamefont {Ali}}, \bibinfo {author} {\bibfnamefont {G.}~\bibnamefont
		{G\"untherodt}}, \bibinfo {author} {\bibfnamefont {B.H.}\ \bibnamefont
		{Hong}}, \bibinfo {author} {\bibfnamefont {B.}~\bibnamefont {Beschoten}}, \
	and\ \bibinfo {author} {\bibfnamefont {B.}~\bibnamefont {\"Ozyilmaz}},\
}\bibfield  {title} {\enquote {\bibinfo {title} {Toward wafer scale
		fabrication of graphene based spin valve devices},}\ }\href@noop {}
{\bibfield  {journal} {\bibinfo  {journal} {Nano Lett.}\ }\textbf {\bibinfo
		{volume} {11}},\ \bibinfo {pages} {2363--2368} (\bibinfo {year}
	{2011})}\BibitemShut {NoStop}%
\bibitem [{\citenamefont {Guimar\~aes}\ \emph {et~al.}(2014)\citenamefont
	{Guimar\~aes}, \citenamefont {Zomer}, \citenamefont {Ingla-Ayn\'es},
	\citenamefont {Brant}, \citenamefont {Tombros},\ and\ \citenamefont {van
		Wees}}]{Guim14}%
\BibitemOpen
\bibfield  {author} {\bibinfo {author} {\bibfnamefont {M.H.D.}\ \bibnamefont
		{Guimar\~aes}}, \bibinfo {author} {\bibfnamefont {P.J.}\ \bibnamefont
		{Zomer}}, \bibinfo {author} {\bibfnamefont {J.}~\bibnamefont
		{Ingla-Ayn\'es}}, \bibinfo {author} {\bibfnamefont {J.C.}\ \bibnamefont
		{Brant}}, \bibinfo {author} {\bibfnamefont {N.}~\bibnamefont {Tombros}}, \
	and\ \bibinfo {author} {\bibfnamefont {B.J.}\ \bibnamefont {van Wees}},\
}\bibfield  {title} {\enquote {\bibinfo {title} {{Controlling spin relaxation
			in hexagonal BN-encapsulated graphene with a transverse electric field}},}\
}\href@noop {} {\bibfield  {journal} {\bibinfo  {journal} {Phys. Rev. Lett.}\
}\textbf {\bibinfo {volume} {113}},\ \bibinfo {pages} {086602} (\bibinfo
{year} {2014})}\BibitemShut {NoStop}%
\bibitem [{\citenamefont {Sydoruk}\ \emph {et~al.}(2014)\citenamefont
	{Sydoruk}, \citenamefont {Go{\ss}}, \citenamefont {Meyer}, \citenamefont
	{Petrychuk}, \citenamefont {Danilchenko}, \citenamefont {Weber},
	\citenamefont {Stampfer}, \citenamefont {Li},\ and\ \citenamefont
	{Vitusevich}}]{Sydoruk14}%
\BibitemOpen
\bibfield  {author} {\bibinfo {author} {\bibfnamefont {V.A.}\ \bibnamefont
		{Sydoruk}}, \bibinfo {author} {\bibfnamefont {K.}~\bibnamefont {Go{\ss}}},
	\bibinfo {author} {\bibfnamefont {C.}~\bibnamefont {Meyer}}, \bibinfo
	{author} {\bibfnamefont {M.V.}\ \bibnamefont {Petrychuk}}, \bibinfo {author}
	{\bibfnamefont {B.A.}\ \bibnamefont {Danilchenko}}, \bibinfo {author}
	{\bibfnamefont {P.}~\bibnamefont {Weber}}, \bibinfo {author} {\bibfnamefont
		{C.}~\bibnamefont {Stampfer}}, \bibinfo {author} {\bibfnamefont
		{J.}~\bibnamefont {Li}}, \ and\ \bibinfo {author} {\bibfnamefont {S.A.s}\
		\bibnamefont {Vitusevich}},\ }\bibfield  {title} {\enquote {\bibinfo {title}
		{Low-frequency noise in individual carbon nanotube field-effect transistors
			with top, side and back gate configurations: effect of gamma irradiation},}\
}\href@noop {} {\bibfield  {journal} {\bibinfo  {journal} {Nanotechnology}\
}\textbf {\bibinfo {volume} {25}},\ \bibinfo {pages} {035703} (\bibinfo
{year} {2014})}\BibitemShut {NoStop}%
\bibitem [{\citenamefont {Kuemmeth}\ \emph {et~al.}(2008)\citenamefont
	{Kuemmeth}, \citenamefont {Ilani}, \citenamefont {Ralph},\ and\ \citenamefont
	{McEuen}}]{Kuemmeth08}%
\BibitemOpen
\bibfield  {author} {\bibinfo {author} {\bibfnamefont {F.}~\bibnamefont
		{Kuemmeth}}, \bibinfo {author} {\bibfnamefont {S.}~\bibnamefont {Ilani}},
	\bibinfo {author} {\bibfnamefont {D.~C.}\ \bibnamefont {Ralph}}, \ and\
	\bibinfo {author} {\bibfnamefont {P.~L.}\ \bibnamefont {McEuen}},\ }\bibfield
{title} {\enquote {\bibinfo {title} {Coupling of spin and orbital motion of
			electrons in carbon nanotubes},}\ }\href@noop {} {\bibfield  {journal}
	{\bibinfo  {journal} {Nature}\ }\textbf {\bibinfo {volume} {452}},\ \bibinfo
	{eid} {06822} (\bibinfo {year} {2008})}\BibitemShut {NoStop}%
\bibitem [{\citenamefont {Viennot}\ \emph {et~al.}({2015})\citenamefont
	{Viennot}, \citenamefont {Dartiailh}, \citenamefont {Cottet},\ and\
	\citenamefont {Kontos}}]{Viennot15}%
\BibitemOpen
\bibfield  {author} {\bibinfo {author} {\bibfnamefont {J.~J.}\ \bibnamefont
		{Viennot}}, \bibinfo {author} {\bibfnamefont {M.~C.}\ \bibnamefont
		{Dartiailh}}, \bibinfo {author} {\bibfnamefont {A.}~\bibnamefont {Cottet}}, \
	and\ \bibinfo {author} {\bibfnamefont {T.}~\bibnamefont {Kontos}},\
}\bibfield  {title} {\enquote {\bibinfo {title} {{Coherent coupling of a
			single spin to microwave cavity photons}},}\ }\href@noop {} {\bibfield
{journal} {\bibinfo  {journal} {{Science}}\ }\textbf {\bibinfo {volume}
	{{349}}},\ \bibinfo {pages} {{408--411}} (\bibinfo {year}
{{2015}})}\BibitemShut {NoStop}%
\bibitem [{\citenamefont {D\'ora}\ \emph {et~al.}(2008)\citenamefont {D\'ora},
	\citenamefont {Gul\'acsi}, \citenamefont {Koltai}, \citenamefont {Z\'olyomi},
	\citenamefont {K\"urti},\ and\ \citenamefont {Simon}}]{Dora08}%
\BibitemOpen
\bibfield  {author} {\bibinfo {author} {\bibfnamefont {B.}~\bibnamefont
		{D\'ora}}, \bibinfo {author} {\bibfnamefont {M.}~\bibnamefont {Gul\'acsi}},
	\bibinfo {author} {\bibfnamefont {J.}~\bibnamefont {Koltai}}, \bibinfo
	{author} {\bibfnamefont {V.}~\bibnamefont {Z\'olyomi}}, \bibinfo {author}
	{\bibfnamefont {J.}~\bibnamefont {K\"urti}}, \ and\ \bibinfo {author}
	{\bibfnamefont {F.}~\bibnamefont {Simon}},\ }\bibfield  {title} {\enquote
	{\bibinfo {title} {Electron spin resonance signal of luttinger liquids and
			single-wall carbon nanotubes},}\ }\href@noop {} {\bibfield  {journal}
	{\bibinfo  {journal} {Phys. Rev. Lett.}\ }\textbf {\bibinfo {volume} {101}},\
	\bibinfo {pages} {106408} (\bibinfo {year} {2008})}\BibitemShut {NoStop}%
\bibitem [{\citenamefont {Horcas}\ \emph {et~al.}(2007)\citenamefont {Horcas},
	\citenamefont {Fern{\'a}ndez}, \citenamefont {G\'{o}mez-Rodr\'{\i}iguez},
	\citenamefont {Colchero}, \citenamefont {G{\'o}mez-Herrero},\ and\
	\citenamefont {Baro}}]{wsxm}%
\BibitemOpen
\bibfield  {author} {\bibinfo {author} {\bibfnamefont {I.}~\bibnamefont
		{Horcas}}, \bibinfo {author} {\bibfnamefont {R.}~\bibnamefont
		{Fern{\'a}ndez}}, \bibinfo {author} {\bibfnamefont {J.M.}\ \bibnamefont
		{G\'{o}mez-Rodr\'{\i}iguez}}, \bibinfo {author} {\bibfnamefont
		{J.}~\bibnamefont {Colchero}}, \bibinfo {author} {\bibfnamefont
		{J.}~\bibnamefont {G{\'o}mez-Herrero}}, \ and\ \bibinfo {author}
	{\bibfnamefont {A.M.}\ \bibnamefont {Baro}},\ }\bibfield  {title} {\enquote
	{\bibinfo {title} {{WSXM: A software for scanning probe microscopy and a tool
				for nanotechnology}},}\ }\href@noop {} {\bibfield  {journal} {\bibinfo
		{journal} {Rev. Sci. Instrum.}\ }\textbf {\bibinfo {volume} {78}},\ \bibinfo
	{pages} {013705} (\bibinfo {year} {2007})}\BibitemShut {NoStop}%
\end{thebibliography}

%

\end{document}